\documentclass[preprint,12pt]{aastex}

\shorttitle{PBRS with CARMA}
\shortauthors{Tobin et al.}

\newcommand{\kms}{\mbox{km s$^{-1}$}}

\newcommand{\lbol}{L$_{\rm bol}$}
\newcommand{\tbol}{T$_{\rm bol}$}
\begin{document}

\title{Characterizing the Youngest Herschel-detected Protostars I. Envelope Structure Revealed by CARMA Dust Continuum Observations}
\author{John J. Tobin\altaffilmark{1,8,9}, Amelia M. Stutz\altaffilmark{2}, S. Thomas Megeath\altaffilmark{3}, William J. Fischer\altaffilmark{3},
Thomas Henning\altaffilmark{2}, Sarah E. Ragan\altaffilmark{2}, Babar Ali\altaffilmark{4}, Thomas Stanke\altaffilmark{5}, P. Manoj\altaffilmark{6},
Nuria Calvet\altaffilmark{7}, Lee Hartmann\altaffilmark{7}}

\begin{abstract}

We present CARMA 2.9 mm dust continuum emission observations of a
sample of 14 Herschel-detected Class 0 protostars in the Orion A and B
molecular clouds, drawn from the PACS Bright Red
Sources (PBRS) sample (Stutz et al.). These objects are characterized by very red 24 \micron\
to 70 \micron\ colors and prominent submillimeter emission, suggesting that they
are very young Class 0 protostars embedded in dense envelopes.  We detect all of
the PBRS in 2.9 mm continuum emission
and emission from 4 protostars and 1 starless core in the fields toward the PBRS;
we also report 1 new PBRS source.
The ratio of 2.9 mm
luminosity to bolometric luminosity is higher by a factor of $\sim$5
on average, compared to other well-studied protostars in the Perseus
and Ophiuchus clouds. The 2.9 mm visibility amplitudes for 6 of the 14 PBRS 
are very flat as a function of uv-distance, with more than 50\% of the source
emission arising from radii $<$ 1500 AU. These flat visibility amplitudes are most
consistent with spherically symmetric
envelope density profiles with $\rho$~$\propto$~R$^{-2.5}$. Alternatively,
there could be a massive unresolved structure 
like a disk or a high-density inner envelope departing from a smooth power-law.
The large amount of mass on scales $<$ 1500 AU (implying high 
average central densities) leads us to suggest that
that the PBRS with flat visibility amplitude profiles are
the youngest PBRS and may be undergoing a brief phase
of high mass infall/accretion and are possibly among the youngest Class 0 protostars.
The PBRS with more rapidly declining visibility amplitudes still have
large envelope masses, but could be slightly more evolved.
\end{abstract}
\altaffiltext{1}{National Radio Astronomy Observatory, Charlottesville, VA 22903}
\altaffiltext{2}{Max-Planck-Institut f\"ur Astronomie, D-69117 Heidelberg, Germany}
\altaffiltext{3}{Ritter Astrophysical Observatory, Department of Physics and Astronomy, University of Toledo, Toledo, OH 43560}
\altaffiltext{4}{NASA \textit{Herschel} Science Center, California Institute of Technology, Pasadena, CA 91125, USA}
\altaffiltext{5}{European Southern Observatory, 85748 Garching bei M\"unchen, Germany}
\altaffiltext{6}{Department of Astronomy and Astrophysics, Tata Institute of Fundamental Research, Colaba, Mumbai 400005, India}
\altaffiltext{7}{Department of Astronomy, University of Michigan, Ann Arbor, MI 48109}
\altaffiltext{8}{Hubble Fellow}
\altaffiltext{9}{Current Address: Leiden Observatory, Leiden University, P.O. Box 9513, 2300-RA Leiden, The Netherlands; tobin@strw.leidenuniv.nl}
\section{Introduction}

Stars form as a result of the gravitational collapse of clouds of gas and dust. This process can take place
in isolated Bok globules \citep[e.g., B335;][]{keene1983,stutz2008,stutz2010} or within fragmented giant molecular
clouds \citep[e.g. Orion;][]{johnstone1999, motte2001, polychroni2013}.
The earliest recognizable phase of the star formation process is the Class 0
phase \citep{andre1993}, the beginning of which is marked by the formation of a hydrostatically
supported protostar within an infalling cloud of gas and dust. 
The robust identification of the youngest sources is imperative for characterizing the initial conditions
at the time of protostar formation, before feedback from the formation process significantly alters local
physical conditions \citep[e.g.,][]{arce2006,offner2014} and probe the earliest stages
of the collapse of the gas onto nascent protostars \citep[e.g.,][]{foster1993}. The
density profile, overall mass, and angular momentum of the 
initially collapsing envelope will determine the potential for fragmentation, 
how quickly the protostar may accumulate mass, and the growth of the circumstellar disk.

At the start of protostellar collapse, just prior to 
protostar formation, there is a theoretical prediction of
 a short-lived ($\sim$1000 yr) first 
hydrostatic core (FHSC), a phase just before or at the start of the Class 0 
phase \citep[e.g.,][]{larson1969,commercon2012}. Several candidate
FHSCs have been identified \citep{enoch2010,chen2010,pineda2011,schnee2012}, but
their identification as true FHSCs remains uncertain, given that these objects could simply be
very low luminosity protostars \citep[e.g., VeLLOs;][]{bourke2006}. 
Since the FHSC phase is thought to be short, 
it is unlikely that many of the candidates in the nearest star forming regions 
(i.e. Perseus, Taurus, Ophiuchus) are true FHSCs due to their relatively small populations of protostars.
Above all else, it is uncertain if there is truly a FHSC phase and if it can be uniquely distinguished
from Class 0 protostars. Nonetheless, detecting and characterizing the youngest 
protostellar sources are key steps towards understanding the
star formation process. To capture short-lived 
phenomena, like the early Class 0 protostars and FHSCs, it is advantageous
to look toward more populous regions of star formation.

The Orion molecular clouds are the nearest regions of active star formation. The \textit{Spitzer} Orion
survey by \citet{megeath2012} found 488 protostellar candidates 
amongst a total of $\sim$3000 young stellar objects.
A subset of 329 protostars from this sample were selected for 
observations in the far-infrared as part of the \textit{Herschel} Orion
Protostar Survey (HOPS) \citep[e.g.,][]{fischer2010,stanke2010,ali2010,manoj2013}. 
Within the fields observed by HOPS, \citet{stutz2013}, hereafter ST13,
serendipitously identified 11 protostars with bright 70 \micron\ and 160 \micron\ emission
that were not part of the original \textit{Spitzer}-selected sample.
At 24 \micron, these sources were either non-detections (8 sources) or so faint that they were
flagged as potential extragalactic contamination in the \textit{Spitzer} surveys \citep{megeath2012,kryukova2012}. Moreover, within the original HOPS sample,
7 protostars had $[24\,\mu{\rm m}]-[70\,\mu{\rm m}]$ colors
 (in log ($\lambda$F$_{\lambda}$) space) redder than 1.65, consistent with 
the \textit{Herschel}-detected sources. ST13 refers to the
18 protostars satisfying the extremely red color criterion as the PACS Bright Red Sources (PBRS).

Analysis of the PBRS spectral energy distributions (SEDs) by ST13, which were augmented by APEX 350 \micron\ and 
870 \micron\ mapping, found that these PBRS sources have 
very cold bolometric temperatures \citep[\tbol;][]{myers1993} (29 K to 45 K)
and high ratios of submillimeter luminosity to bolometric luminosity (\lbol) 
(0.6\% to 6.1\%). Most PBRS are not
detected shortward of 24 \micron, but some display faint 
features in the \textit{Spitzer} IRAC 4.5 \micron\
band, possibly indicative of shocked H$_{2}$ emission associated with outflows. 
Despite their deeply embedded nature, sources
emitting at 70 \micron\ must be self-luminous. For example, starless core models 
show that the emission would otherwise be too faint to detect at 
70 \micron\ \citep[][ST13]{ragan2012}. It is important to point 
out that the PBRS sources are \textit{not} low-luminosity objects like the VeLLOs \citep[e.g.,][]{bourke2006} since
they have \lbol\ ranging between 0.65 L$_{\sun}$ 
and 30.6 L$_{\sun}$, with the median being $\sim$3 L$_{\sun}$. These luminosities
are large enough such that they are not dominated by external heating \citep{dunham2008}.

The characteristics of the PBRS indicate that these protostars could be very young
Class 0 sources with very dense envelopes (ST13). There is, however, a degeneracy in the 
interpretation of protostar SEDs between envelope density and
inclination due to bipolar cavities being evacuated by the outflows. The envelope properties were
also difficult to study with only the APEX submillimeter (submm) data available in ST13, due to the low resolution 
and blending of the envelope emission with extended cloud structure. Furthermore, the lack 
emission shortward of 10 \micron\ toward most PBRS, made the inclinations impossible to constrain
from SED modeling. The only way to derive more
detailed envelope properties, independent of inclination, is to observe these sources with an interferometer.

We have obtained observations of a subset of the PBRS sample with the Combined Array for Research 
in Millimeter-wave Astronomy (CARMA). There are a total of 
19 PBRS, 18 of which were presented in ST13 and 1 additional
PBRS will be described in this paper. We
have observed
14 PBRS, focusing on the new, \textit{Herschel}-detected subset of
PBRS. We focused on this subset because they had the least amount of complementary data and 
non-detections at most wavelengths shorter than 70 \micron. 
We observed the PBRS in both the dust
continuum and spectral line emission to examine both the envelope and outflow 
properties of these sources; the outflow results will be presented in a future paper.
We discuss
the observations in Section 2, our results for the dust continuum emission and 
model comparison are given in Section 3, we discuss the results within the 
broader context of star formation in Section 4, 
and summarize our main conclusions in Section 5.

\section{CARMA Observations and Data Reduction}

CARMA is a heterogeneous interferometer array comprised of 23 antennas (6 - 10.4 m,
9 - 6.1 m, and 8 - 3.5 m) located in the Inyo mountains of California. Our
observations were carried out with the main, 15-element CARMA array using
the 10.4 m and 6.1 m antennas in two configurations. 
We observed a subset of the PBRS identified in ST13 
with CARMA in the D configuration during late 2012 and early 2014. 
We also followed-up a subset of the sources observed in 
D-configuration with higher-resolution observations in C-configuration in early 2014. 
The angular resolutions
in D and C configurations were $\sim$5\arcsec\ and $\sim$2\arcsec\ respectively.
The central frequency was 107.77 GHz and four spectral windows were configured 
for 500 MHz bandwidth to observe the dust continuum; two windows
were configured for 8 MHz bandwidth to observe para-NH$_2$D ($J=1_{11}\rightarrow1_{01}$) and 
C$^{18}$O ($J=1\rightarrow~0$); and the two remaining windows had 31 MHz 
bandwidth to observe $^{13}$CO ($J=1\rightarrow0$) and $^{12}$CO ($J=1\rightarrow0$).
Two or three sources were observed per track, with further details given in Table 1. 
The C-configuration observations did not observe para-NH$_2$D and another
500 MHz continuum band was allocated. Generally, 
we only detect $^{12}$CO ($J=1\rightarrow0$) and the 2.9 mm continuum; 
there were a few weak detections in the other lines which will not 
be discussed further. Our root-mean-squared (RMS)
sensitivity is typically 0.2 Jy beam$^{-1}$ channel $^{-1}$ 
for the CO ($J=1\rightarrow0$) and 1 mJy beam$^{-1}$ for the continuum data.
The data were reduced, edited, and imaged using standard procedures within the MIRIAD 
software package \citep{sault1995}. The uncertainty in the absolute flux
is estimated to be $\sim$20\%. 
We will only present the continuum results in this paper, the CO outflow data will be 
presented in an upcoming paper.

\section{Results}

The observations of $\lambda$ = 2.9 mm continuum emission
enable us to probe the properties of the protostellar envelopes, in
terms of mass and density profiles. We will discuss the overall flux densities, visibility
amplitudes, and comparison of the visibility amplitudes to radiative transfer models.
We also report the data observed toward additional sources located within our field of view,
 but primarily discuss the PBRS in the main text; the PBRS and non-PBRS are denoted in Table 2. 

\subsection{Integrated 2.9 mm Dust Continuum Emission}

We detect all the observed PBRS sources in the 2.9 mm continuum and deconvolved images
using natural weighting are shown in Figure 1. 
Our observations are sensitive to spatial scales between $\sim$1000 AU and $\sim$10000 AU.
On these scales, most sources have some resolved structure, in terms of extended envelope emission
and in the case of HOPS 373, there is a binary source separated by $\sim$4\arcsec. 
Two sources (082012 and 061012) also have companions $\sim$20\arcsec\ (9400 AU) away. 
Images of the sources made with Robust weighting factor of -1 \citep{briggs1995} 
did not reveal significant structure on smaller-scales.
The flux densities measured from the deconvolved images 
are presented in Table 2. 
The 2.9 mm flux densities of the sample exhibit a relatively
large amount of heterogeneity given the extremely red colors selection of the sample.
The brightest PBRS is 082012 at 155.6 mJy and the faintest is 119019 at 10.2 mJy. Indeed, 12 of 14 PBRS
have flux densities $>$ 30 mJy and their values of \lbol\ also span an order of magnitude. 
The combined D- and C-configuration images agree with D-configuration-only flux
densities within the statistical uncertainties. 
Note that we also present an additional
PBRS, 135003, that did not appear in ST13. This source was left out from the sample
due to the 70 \micron\ FWHM being extended more than the cutoff value 7\farcs8. More details of this source
and its infrared and submillimeter imaging are given in the Appendix; its inclusion
raises the number of PBRS in the Orion clouds to 19.

The strength of dust continuum emission from the PBRS sources prompted us to
 collect $\lambda$ $\sim$ 3 mm flux densities from the literature of other 
Class 0 or Class I protostellar sources observed with interferometers 
for comparison \citep[Table 3; ][]{looney2000,arce2006,tobin2011}).
These observations had comparable resolution and sampling of the uv-plane.
To match the 2.9 mm flux densities better, we have scaled the flux densities of the comparison sources.
We have done the scaling by assuming that the relative flux densities only depend on the the dust
opacity spectral index ($\beta$) and the 
function F$_{\lambda}$ $\propto$ $\lambda^{-(2 + \beta)}$. This assumption is reasonable
given the similar wavelengths of the samples. With the further assumption that $\beta$ $\sim$ 1,
the scaling factors for the 2.7 mm flux densities and 3.4 mm flux densities
are 0.8 and 1.6 respectively.
We have converted all the flux densities to 2.9 mm luminosities (L$_{2.9mm}$) using 
the distances provided in Table 3
and assuming a bandwidth of 0.11 mm (4 GHz).
We plot L$_{2.9mm}$ versus \lbol\ and 
the ratio of L$_{2.9mm}$ to \lbol\ versus \lbol\ for all the data in Figure \ref{mmflux}. 

The comparison sources span the range of observed \lbol\ for the PBRS, but
most PBRS have lower \lbol\ values, comparable to those in \citet{tobin2011}.
They, however, have L$_{2.9mm}$ values
that are comparable to the \citet{looney2000}
sources, which are among the brightest nearby protostars a millimeter wavelengths 
(e.g. NGC 1333 IRAS 4A, NGC 1333 IRAS2A, IRAS 16293-2422) and are more
luminous than most PBRS. Furthermore, the PBRS
have among the largest values of L$_{2.9mm}$ and L$_{2.9mm}$/\lbol\ ratios. 
This behavior is true at all luminosities, but especially evident at \lbol\ $\sim$ 1 L$_{\sun}$.
The non-PBRS sources in our observations generally have higher \lbol, higher T$_{\rm bol}$, and lower L$_{2.9mm}$/\lbol\ ratios;
however, the source HOPS 68 does intermingle with the PBRS in Figure \ref{mmflux}. Note that the results do not significantly change 
whether or not scaling is applied to the literature data.

Also evident in Figure \ref{mmflux} is the lack of a clear 
relationship between \lbol\ and L$_{2.9mm}$. This indicates that the millimeter
emission is decoupled from the central source
properties (central source refers to both the protostar and accretion processes generating
luminosity). The PBRS are tracing a new region of parameter space with their
large amounts of circumstellar material traced by the 2.9 mm flux densities and lower values
of \lbol.

\subsection{Circumstellar Masses}

The integrated flux densities of the protostars enable us to directly probe
the circumstellar mass associated with the protostars, without significant
contributions from the surrounding molecular cloud. In this case, the interferometer
filtering works to our advantage by separating the envelope emission from
the surrounding background cloud. To convert a flux density into a mass, we assume that the emission is optically thin
and isothermal, and apply the equation
\begin{equation}
M = \frac{D^2 F_{\lambda} }{ \kappa_{2.9mm} B_{\lambda}(T_{dust}) };
\end{equation}
where $B_{\lambda}$ is the Planck function. We have assumed that $T_{dust}$ = 20 K, 
$\kappa_{2.9mm}$ = 0.00215 cm$^{2}$~g$^{-1}$ using \citet{ossenkopf1994} 
(Table 1, Column 5) extrapolated to 2.9 mm, and $D$ = 420 pc.
The extrapolation to 2.9 mm uses the dust opacity spectral index ($\beta$) of the \citet{ossenkopf1994} 
dust model between 700 \micron\ to 1.3 mm which has $\beta$ = 1.78. The opacity given 
is the dust+gas opacity, assuming a gas-to-dust ratio of 100.

The calculated masses are given in Table 2, the uncertainties given are statistical only (not
including the uncertainty in absolute flux calibration) 
and the masses themselves are likely only valid at the order of magnitude
level given the assumptions. There may be optically thick regions of the envelope, but those are on scales of
order a few hundred AU and will make only a small contribution to the overall mass.

With these assumptions,
all the PBRS sources (except 119019) have more 
than 1 $M_{\sun}$ of surrounding material, with the largest
being almost 10 $M_{\sun}$. These masses are reflected in the high integrated flux densities observed toward these protostars.
We also note that the masses are systematically larger than those calculated by ST13.
The masses calculated in ST13, however, are from modified blackbody fits to the emission from 70 \micron\ to 870 \micron\ and
the ST13 modified blackbody fits
systematically underpredict the 870 \micron\ flux densities. The underprediction of the 870 \micron\ flux densities likely
results from fitting a single temperature to data that reflect a superposition of temperatures and
span an order of magnitude in wavelength. If masses were calculated
directly using the 870 \micron\ flux, closer agreement is expected. Several
non-PBRS have masses listed in Table 2 that are comparable to the mass 
of the PBRS. Many of these sources, however, have higher \lbol\ and T$_{\rm bol}$,
suggesting that the dust temperatures could be larger and by extension the masses
are overestimated. We assumed T$_{dust}$ = 20 K,
 and the actual masses will be different by the ratio T$_{dust}$/20~K.

The estimated masses will also change if we assume
a different dust opacity law; \citet{ossenkopf1994} has $\beta$ = 1.78 at millimeter
wavelengths (Table 1, Column 5). If we instead assume $\beta$ = 1 and use the normalization of 
0.1($\nu$/1200 GHz)$^{\beta}$ from \citet{beckwith1990},
the masses would be a factor of 4.5 lower. 

\subsection{Visibility Amplitudes of 2.9 mm Continuum}

The integrated flux densities are only one aspect of the continuum data,
the visibility amplitudes as a function of uv-distance/baseline length can reveal
more about the source structure than the deconvolved images alone. We show the visibility amplitudes
for all detected sources in Figure \ref{uvamps}. 
How slowly (or rapidly) the amplitude decreases with increasing uv-distance reveals 
how concentrated the emission is toward a particular source, in addition to structural changes in the emitting
material.
 Similar to the order of magnitude span in 2.9 mm flux density, the
visibility amplitudes profiles themselves are quite varied but generally fall within two groups. About half of 
the observed PBRS have amplitudes that drop quickly with 
increasing uv-distance (HOPS 373, 082012, 302002, 119019, HOPS 372, 019003A, 061012), 
meaning that there is more emission on larger spatial scales relative to small spatial scales.
The other half of the sample have amplitudes that are flat or slowly decreasing 
with increasing uv-distance (093005, 090003, 091016, 097002), indicative
of most emission arising from compact, unresolved structure.
 The visibility amplitudes of 082005 and 091015
are most consistent with the flat visibility amplitude sources, 
but decrease more rapidly than the others.
We note that 135003 was only observed in D-configuration and is located in a more complex region, 
so it is uncertain if its flat visibilities extend toward larger uv-distances.

To examine the scales at which most flux is being emitted, we plot the ratio of the visibility amplitudes
at 5 k$\lambda$ ($\sim$41\arcsec, 17300 AU) F(5k$\lambda$) to those at
30 k$\lambda$ ($\sim$7\arcsec, 3000 AU) against F(30k$\lambda$) and \lbol\
in Figure \ref{uvratio}. We see that the brightest PBRS sources at 30 k$\lambda$ also have the lowest ratios, meaning that
most of their flux is emitted from scales smaller than 3000 AU; 082012 is an outlier from this trend.
When plotted against \lbol, the F(5k$\lambda$)/F(30k$\lambda$) ratio tends to be $<$ 2 for sources
with luminosities of $\sim$1 \lbol, while
higher luminosity sources and non-PBRS tend to have ratios $>$ 2.  We note, 
however, that a source composed of just a circumstellar disk would 
appear to have a ratio of 1 on these plots and the non-PBRS sources
that have F(5k$\lambda$)/F(30k$\lambda$) ratios $\sim$ 2 are likely 
more-evolved protostars whose millimeter emission is likely to be dominated by a disk on
small spatial scales.

\subsection{Comparison to Protostellar Envelope Models}

The visibility amplitude profiles can also indicate the density profiles of the envelope.
We ran a small grid of radiative transfer models to obtain qualitative results for the
interpretation of the visibility amplitude data. The goal is to determine what 
density profiles are consistent with the data and if a compact, unresolved
source is a necessary component for the models to fit the data. 
We use the Hyperion code \citep{robitaille2011} to perform the radiative equilibrium calculations
and produce ray-traced images of 2.9 mm continuum emission, with a 5.0 $M_{\sun}$ 
envelope, 1 L$_{\sun}$ central protostar, 10000 AU outer radius, and radial density profiles
of $\rho$ $\propto$ R$^{-1.5,-2.0,-2.5}$, and a 50 AU radius 
embedded disk with $M_{disk}$ = 0.0 $M_{\sun}$, 0.01 $M_{\sun}$, and 0.1 $M_{\sun}$. We also ran envelope models using
the density structure for a rotationally-flattened, infalling envelope 
\citep[CMU envelope;][]{ulrich1976, cassen1981}.
For the CMU models, we explored the same disk masses, 
but we used four centrifugal radii ($R_C$ = 50 AU, 100 AU, 300 AU, and 500 AU) and
assumed that the disk radius was equal to $R_C$; $R_C$ is the radius at which infalling material can
be rotationally-supported due to conservation of angular momentum. 
The overall envelope masses of the CMU
models were the same as those of the power-law envelopes. The inclination of the system only 
has a minor effect on the visibility amplitudes and, we assume an inclination 
angle of 60\degr\ for simplicity.

We use the dust opacities calculated by \citet{ormel2011} for icy silicate grains and 
bare graphite grains grown for a period of 3$\times$10$^5$ yr. These dust opacities 
are similar to those of \citet{ossenkopf1994} (Table 1, Column 5), but are calculated 
down to $\lambda$ $\sim$0.1 \micron\ and include scattering properties. The dust opacity spectral
index ($\beta$) of the \citet{ormel2011} models, however, is $\sim$2 at submillimeter and millimeter wavelengths, 
consistent with ISM-sized dust grains. This
steep $\beta$,
however, results in a very 
low dust opacity at 2.9 mm and very faint envelope emission. Therefore,
we have altered the dust opacity model and at wavelengths greater 
than 90 \micron\ we transition to the \citet[][Table 1 Column 5]{ossenkopf1994} dust opacity 
model. This model has  $\beta$ $\sim$ 1.78, yielding $\kappa_{2.9mm}$ = 0.00215, producing 
2.9 mm millimeter fluxes more consistent with our observations.

We could have simply increased the envelope masses such that the flux densities were consistent with our data.
The \citet{ormel2011} dust opacities, however, are a factor of 2.35 lower 
than \citet[][Table 1 Column 5]{ossenkopf1994}. Thus, it would
have been necessary to increase envelope masses to $\sim$12 $M_{\sun}$, which 
may be unrealistically large for many of our sources. Furthermore,
the larger masses would increase the envelope opacity at shorter wavelengths 
and make the overall dust temperatures lower. There is evidence for
millimeter-sized dust grains in protostellar envelopes which would cause a 
shallower $\beta$ of $\sim$1 \citep{sadavoy2013,kwon2009,schnee2014} and we
therefore feel justified in adopting a hybrid dust opacity model.

We generated 30000 AU $\times$ 
30000 AU model images  with 15 AU resolution, corresponding to 2048$\times$2048 pixel images for emission 
between 2.8 mm and 3.0 mm.
Such high resolution was necessary to ensure that we did not introduce false structure
when Fourier transforming the images to compare with the observed visibility data. We used the 
MIRIAD task \textit{fft} to calculate the Fourier transform of each model image and
we azimuthally averaged the Fourier transformed image to construct a 1-dimensional visibility amplitude
profile. To facilitate model comparisons, we normalized the flux densities of the envelopes and data 
in our comparisons at uv-distances of 
22.5 k$\lambda$. This normalization is reasonable because most 
emission should be optically thin at $\lambda$ $\sim$ 3 mm on spatial scales 
larger than our best resolution ($\sim$2000 AU) and we are only interested
in comparing the density profiles, not fitting model envelopes in detail.

We perform a simple model comparison for the sources with the best 
signal-to-noise ratios at the uv-distances probed by our observations: 
082005, 093005, 090003, 082012, and 097002, see Figures \ref{uvcomps} and \ref{uvcomps-cmu}. 
This sample  is representative of the range in visibility amplitudes profiles observed, e.g., from
very flat (090003, 097002, and 093005), intermediate (082005), and
rapidly declining (082012).

The flat visibility amplitude sources (090003, 097002, and 093005) are
consistent with a power-law envelope with $\rho$ $\propto$ R$^{-2.5}$ 
envelope if there is no unresolved component.
When a 0.01 $M_{\sun}$ unresolved component is included in a power-law envelope,
the flat visibility sources are still most consistent with a $\rho$ 
$\propto$ R$^{-2.5}$ envelope. We note that the observed visibility amplitude profiles of 
090003, 097002, and 093005 are systematically elevated with respect to 
the $\rho$ $\propto$ R$^{-2.5}$ envelope both with and without the inclusion of a 0.01 $M_{\sun}$ disk.
When comparing the flat visibility sources to models with a 
0.1 $M_{\sun}$ unresolved component, all three power-law envelope 
density models provide a reasonable match to the visibility data since the compact 
source dominates the emission at uv-distances $>$ 20 k$\lambda$, although the curvature of visibility 
curve of the R$^{-1.5}$ model is opposite to that apparent in the data. In the case of 
the CMU envelope, the flat visibility amplitude sources 
are \textit{inconsistent} with no disk component and are \textit{consistent} with a 0.1 $M_{\sun}$ disk
with $R_C$ $\le$ 100 AU, but the curvature of the model visibility amplitude profile
 is in the opposite sense as the data.

The intermediate source between the extremes of flat visibilities and rapidly declining visibilities
(082005) is consistent with a density profile between $\rho$ $\propto$ R$^{-2.0}$ and $\rho$ $\propto$ R$^{-2.5}$
if no unresolved component is included. It is consistent with $\rho$ $\propto$ R$^{-2.0}$ when a  
0.01 $M_{\sun}$ unresolved component is included, but is marginally inconsistent with a power-law envelope
and a 0.1 $M_{\sun}$ unresolved component. Assuming a CMU envelope, it is most consistent with
$R_C$ = 300 AU and a 0.1 $M_{\sun}$ disk.

The rapidly declining visibility amplitude source 082012 is well-matched
by the power-law envelope models with no unresolved component and a density profile between 
$\rho$ $\propto$ R$^{-2.0}$ and $\rho$ $\propto$ R$^{-1.5}$. The data are also consistent 
with $\rho$ $\propto$ R$^{-1.5}$ when a 0.01 $M_{\sun}$ 
disk is included. If we assume a CMU envelope structure, then 
082012 is also consistent with a CMU envelope with $R_C$ = 100 - 300 AU, containing 
a 0.01 $M_{\sun}$ disk. Thus, sources with rapidly declining
visibility amplitudes (082012 and others in the sample)
are \textit{inconsistent} with both density profiles steeper than $\rho$ $\propto$ R$^{-2}$ and
disk components more massive than 0.01 $M_{\sun}$.

In addition to comparing the visibility amplitude profiles directly, we show the visibility
amplitude ratios from the envelope-only models in Figure \ref{uvratio}. These ratios can be thought of as
limiting cases in the absence of a massive protostellar disk. The visibility amplitude ratio
for the $\rho$ $\propto$ R$^{-2.5}$ envelope is the smallest, but it is still in excess of observed sources with the smallest
ratios. Most observed sources have visibility amplitude ratios in between the values
found for the $\rho$ $\propto$ R$^{-2.5}$ and $\rho$ $\propto$ R$^{-2.0}$ models. The addition of an unresolved
component to any of the models in Figure \ref{uvratio} would decrease the ratios, making the models more consistent
with the observations, but with a shallower density profile.

The qualitative model comparison shows that multiple physical structures can
be invoked to explain both the flat and rapidly declining visibility amplitude
sources. The flat visibility sources can be explained with having most 
flux in the unresolved component or a very steep ($\rho$ $\propto$ R$^{-2.5}$) density profile. 
Thus, a steep density profile 
is essentially indistinguishable from a compact source with our current data.
The rapidly declining visibility sources, on the other hand,
are inconsistent with a massive unresolved 
component (0.1 $M_{\sun}$) within a power-law or CMU envelope.
There may, however, be some additional dependence on the disk density structure
that we do not explore here.
Higher resolution data will be necessary to break these degeneracies between
power-law envelopes with no or an unresolved component and rotationally-flattened
envelopes with a large, massive disk component.

\section{Discussion}

The PBRS represent an intriguing piece to the puzzle of low-mass 
star formation. Their 2.9 mm luminosities are quite large relative to their
bolometric luminosities, and their millimeter luminosities are comparable to the 
brightest millimeter sources known in the nearby star forming regions. At the same time
4 out of the 14 observed PBRS  have some of the flattest 2.9 mm 
visibility amplitudes observed toward any protostellar source; indeed, the most comparably flat source
is NGC 1333 IRAS 4B \citep{looney2003}. In the following subsections,
we compare and contrast the PBRS to known protostars in nearby star forming regions and 
theoretical models to examine their significance in the star formation process as a whole.

\subsection{Envelope Density Profiles}

The favored interpretation of the the very red 24 \micron\ to 70 \micron\ colors exhibited by 
the PBRS is very high envelope densities (ST13). Most of the observed PBRS
envelopes appear to be quite massive as measured from their 2.9 mm flux densities (Table 2).
Furthermore, the low 5 k$\lambda$ to 30 k$\lambda$ flux ratios
indicate that there is a significant amount of unresolved emission 
at spatial scales less than 3000 AU (in diameter)
for 6 of 14 sources. Two possible explanations
for the flat visibility amplitudes are either steep envelope density profiles or massive--compact structures with
densities in excess of a smooth power-law density profile.
Analytic protostellar collapse models predict several radial 
density profiles that we could expect to observe. 

The Larson-Penston solution \citep{larson1969} (and the numerical solution)
predicts that the free-fall collapse of a constant
density cloud would result in a $\rho$ $\propto$ R$^{-2}$ density profile. 
Bonnor-Ebert spheres \citep{bonnor1956} on the other hand, have a high, 
constant density region, with a surrounding envelope with a 
$\rho$ $\propto$ R$^{-2}$ density profile. As a Bonnor-Ebert 
sphere collapses, the entire density profile approaches $\rho$ $\propto$ R$^{-2}$. 
A Bonnor-Ebert sphere with a small, flat inner region cannot account for the observed density
structures since the density of the inner region joins smoothly with the outer power-law envelope
and our observations would require a jump to higher density. Moreover, since all the sources
we observe are protostellar, we do not expect there to be a flat density region at small 
spatial scales.

A singular isothermal sphere (SIS) also has a $\rho$ $\propto$ R$^{-2}$ density profile and
this is the initial condition of the \citep{shu1977} protostellar
collapse model. The free-fall collapse of a SIS is inside-out, meaning there is an 
outwardly propagating rarefaction wave that bounds the infalling region of the envelope;
the infalling region has a $\rho$ $\propto$ R$^{-1.5}$ density profile. 
This model was extended to include rotation by \citet{tsc1984} and 
the region inside the centrifugally supported radius has a density profile 
of $\rho$ $\propto$ R$^{-0.5}$. Within the context of these models, the envelope emission
of the youngest sources is expected to be dominated by the $\rho$ $\propto$ R$^{-2}$ region
since both the infall and rotationally supported regions are small at early times.
For an initial sound speed (c$_s$) of 0.2 \kms\ (assuming T = 10 K),
the infalling region would 
extend to a radius of $\sim$1050 AU (2\farcs5) 25 kyr after collapse begins (r = $c_s$ $\times$ t). Thus, 
for extremely young sources, one would expect that data with a maximum resolution of $\sim$2\arcsec\
to be dominated by the $\rho$ $\propto$ R$^{-2}$ component of the envelope (within the context of this model).
If some of the sources, however, are more evolved, with larger collapsing regions, the shallower density
profiles and possibly the rotationally-flattened portion
of the density structure would be apparent in the visibility amplitudes (see Figure \ref{uvcomps-cmu}).

The envelope models that we compare to the observed visibility amplitudes
in Figure \ref{uvcomps} have density profiles that encompass those
expected from the
theories of protostellar collapse. Considering \textit{only} the envelope
contribution to the visibility amplitudes \textit{without} additional unresolved 
source emission, however, the sources with 
flat visibility amplitudes (090003, 093005, 097002, 135003
and 091016) models are most consistent with the envelope density profiles
that are as steep as $\rho$ $\propto$ R$^{-2.5}$. 
The sources with more rapidly declining
visibility amplitudes (082012 and 302002) are consistent with 
the $\rho$ $\propto$ R$^{-1.5}$ density profile.
The sources in between (082005 and 091015) are consistent with  $\rho$ $\propto$ R$^{-2}$
density profiles. Thus, if we consider only envelopes with smooth radial density
profiles, the flat visibility amplitude sources appear to be
inconsistent with the analytic protostellar collapse theories, while the sources with intermediate and rapidly
declining visibility amplitudes fall within theoretical expectations.

Such steep density profiles
have been obtained previously toward more nearby protostars \citep{looney2003,chiang2008,kwon2009,chiang2012}. 
The sources with flat visibility amplitudes appear similar to 
NGC 1333 IRAS 4B, a source with very flat visibility 
amplitudes out to $\sim$80 k$\lambda$ \citep{looney2003,chiang2008}; 
this is equivalent to flat visibility amplitudes out to 
$\sim$150 k$\lambda$ at the distance to Orion.
The other sources with more rapidly declining visibility amplitudes
appear more similar to NGC 1333 IRAS 4A, or IRAS 2A.
Our results further confirm that the dust continuum structure of some protostellar envelopes 
indicate radial density profiles steeper than expected from analytic
models of collapse, consistent with previous findings by \citet{looney2003} and \citet{kwon2009}.

The density profiles steeper than the analytic models could result from having 
asymmetric envelope structures on these scales. \citet{tobin2010,tobin2012} showed that
envelopes are often filamentary and asymmetric on $>$1000 AU scales and that infall might 
come through a filamentary envelope rather than a spherical envelope. The \textit{Spitzer} 
IRAC and the APEX 350 \micron\ and 870 \micron\ images in ST13 often show 
filamentary structure on larger scales that may persist on
smaller scales (e.g., Figures 8 and 13b of ST13).
The 2.9 mm continuum images described here, however, do not have features suggestive of 
strong asymmetry in most cases, but \citet{tobin2010} argued 
that asymmetry can be difficult to observe in the dust continuum
due to the emission resulting from a combination of density and temperature.

If additional, unresolved components to the dust emission are considered (i.e., added to the
envelope density profile), the flat visibility amplitude sources may be 
consistent with shallower density profiles. With an unresolved component
of 0.1 $M_{\sun}$, then the $\rho$ $\propto$ R$^{-2.0}$ 
and R$^{-1.5}$ density profiles (in addition to $\rho$ $\propto$ R$^{-2.5}$) 
are able to reproduce the flat visibility amplitudes observed for some sources
(Figure \ref{uvcomps}). Even with the unresolved component, however, the curvature in the visibility amplitudes
of the $\rho$ $\propto$ R$^{-1.5}$ density profile is inconsistent with the data. The steeper density
profiles also do not fully capture the curvature observed in the data, but the effect
is less dramatic than for $\rho$ $\propto$ R$^{-1.5}$.

A circumstellar disk is a natural compact structure that is expected to develop during the 
star formation process \citep[e.g.,][and references therein]{williams2011} and are
likely to have a variety of sizes \citep{maury2010,tobin2012}. Therefore, a comparison to the CMU models
with and without disk components is also of interest, see Figure \ref{uvcomps-cmu}). The
 rapidly declining and intermediate sources could be consistent with having a large $R_C$
region (and a large protostellar disk), depending on the disk mass. The
sources with flat visibility amplitudes can only be consistent with small $R_C$ and a massive disk.
Again, however, the curvature of the model visibility amplitude curves with small (R = 50 AU, 100 AU), 0.1 $M_{\sun}$ disks 
are dissimilar to that of the data. Thus, the models show that disk components are possible in the intermediate and rapidly
declining cases, but the exact parameters are not well-constrained. The flat visibility amplitude sources on the
other hand are grossly consistent with a disk component within the context of the CMU model,
 but, as stated previously, the curvature of the visibility amplitude profiles of the models
relative to the observations are different.

Rather than a massive disk component, it is also possible that the envelope 
density profile itself is not a smooth power-law. 
Simulations of protostellar collapse including 
magnetic fields have been shown to 
create density enhancements of infalling material during collapse 
that depart from a smooth power-law density profile
\citep{tassis2005}. Such structures can 
cause flattening of the visibility amplitudes due to the mass build-up
at small spatial scales. The robustness of this model, however,
is unclear.

Regardless of the exact nature of the structure, we can conclude that the envelopes with
flat visibility amplitudes are 
inconsistent with the often assumed $\rho$ $\propto$ R$^{-1.5}$ density profile for
protostellar envelopes. The flat visibility amplitude profiles are most consistent with
either a
steep density profile ($\rho$ $\propto$ R$^{-2.5}$)
or the visibility amplitudes are dominated by dense, unresolved structure. The unresolved structure
could be a disk, or it could be departures from a power-law density profile. Higher 
resolution data that resolve down to the expected scales of a disk are necessary 
to distinguish between these two (radically different) scenarios.

\subsection{Nature of the PBRS}
The defining characteristic of the PBRS from the \textit{Herschel}
study by ST13 is the very red color of the PBRS sample as a whole,
having $[24\,\mu{\rm m}]-[70\,\mu{\rm m}]$ colors
 (in log ($\lambda$F$_{\lambda}$) space) redder than 1.65. Furthermore,
the \tbol\ measurements of the PBRS are in a narrow range of 20 - 45 K, consistent
with little observed emission shortward of
24 \micron\ for most PBRS. Thus, while the 
coldness (and redness) of the SEDs of the PBRS is consistent throughout the sample, they
are very heterogeneous in terms of their ratio of L$_{submm}$ to \lbol\
 (0.6\% to 6.1\%), \lbol\ itself, and 2.9 mm luminosity.  
Nevertheless, the PBRS are characterized by higher L$_{2.9mm}$ 
to \lbol\ ratios than previously identified in protostellar samples (see Figure \ref{mmflux}).

The expected evolutionary trend for protostars is that they become more luminous as they accrete mass due to
increased photospheric luminosity and greater accretion luminosity \citep{young2005, dunham2010}.
\tbol\ is also expected to also increase with decreasing envelope density and 
hence optical depth. At the same time, clearing of the envelope 
is likely driven by the influence of the protostellar outflow \citep{arce2006,offner2014}.
Thus, it is expected that very young Class 0 protostars will have larger millimeter 
flux densities (or larger envelope mass), with rather low luminosities; in other words they will have
large fractions of millimeter luminosity relative to \lbol\ \citep[e.g.,][]{andre1993}. 
These changes, however, may not be due solely to evolution: both initial conditions and evolution
play roles in the observed \tbol\ and 
L$_{submm}$/\lbol\ ratios observed toward particular protostars \citep[e.g.][]{young2005}.
Further complicating matters, \tbol\ can be strongly influenced by the 
orientation of a given source in the plane of the sky, such that a more 
evolved source viewed edge-on can appear younger 
\citep[ST13; ][]{jorgensen2009,launhardt2013,dunham2014}. 
Within the sample of PBRS, \tbol\ is confined
to a narrow range, but there are a few sources that have 
low luminosities and low 2.9 mm flux densities (061012 and 119019). 
Meanwhile, other PBRS have both relatively high luminosities and high millimeter flux densities
 (082012 and HOPS 373). Such variations in the observed properties of the sample 
indicate that, despite the stringent color selection (ST13), the PBRS as a 
whole may not be characterized by a single evolutionary state.  
Furthermore, the fact that the PBRS present such a narrow range 
in \tbol\ but exhibit a broad range in other properties poses 
possible problems for a \tbol-based classification of protostars. 
If even at the lowest values of \tbol\ and for a uniformly selected 
sample we see clear variations, then even larger variations may be seen across the \tbol\
range encompassing the Class 0 and Class I protostellar phases. Thus, not all
protostars of equivalent \tbol\ are equal.

With the interferometry data, we are able 
to determine the spatial scales from which we are detecting emission due to the
high-resolution and analysis of visibility amplitude profiles.
The visibility amplitude profiles of the PBRS have many features, but
they can be broadly described as flat or rapidly declining. The visibility amplitude
ratios from 5 k$\lambda$ to 30 k$\lambda$ (flux at $\sim$17,000 AU to 3000 AU scales) plotted versus 
30 k$\lambda$ flux density and \lbol\ (see Figure \ref{uvratio}), further enable the sample
to be examined as a whole. The expected evolutionary trends
for visibility amplitude ratios and profiles are uncertain due to the unknown 
contribution of the disk at a given time. Nevertheless, we can use the analytic models
for protostellar envelopes as limiting cases.

 If we consider the collapse of a 
singular isothermal sphere with a density profile $\propto$ R$^{-2}$, the visibility amplitude
ratios would be smaller initially and then increase as material falls in from
larger radii. The density profile of the infalling region 
will be proportional to R$^{-1.5}$ and this region grows with time.
So, in the absence of a massive disk, the visibility amplitude ratio for such a model
will be between the R$^{-2}$ and R$^{-1.5}$ values (see ratios taken from the 
models in Figure \ref{uvratio}). Then as a disk grows and
the envelope dissipates, the 5 k$\lambda$ to 30 k$\lambda$ visibility amplitude ratios
will decrease (trending toward 1 on the scales examined)
as the disk begins to dominate the flux density of the system at millimeter wavelengths. Thus, to summarize
the visibility amplitude ratios will start small, then increase, and then decrease. 
The exact evolution will depend on how quickly a large disk forms and
how massive such a disk is. 
We note that if a singular isothermal sphere formed from a Bonnor-Ebert sphere, a Bonnor-Ebert sphere
would initially have a large visibility amplitude ratio, depending on the size of the flat density region \citep[e.g.,][]{schnee2012}.
The visibility amplitude ratios would then decrease as the Bonnor-Ebert sphere evolved toward a singular isothermal sphere.

Two trends are evident in the visibility amplitude ratio plots shown in Figure \ref{uvratio}. 
Most of the sources with the highest flux densities at 30 k$\lambda$ also have the lowest 5 k$\lambda$ to
30 k$\lambda$ ratios (i.e., spatially compact flux density). Then 
looking at the 5 k$\lambda$ to 30 k$\lambda$ ratio versus \lbol\, we see that 6 out of 14 
of the lower luminosity PBRS sources (\lbol\ $<$ 3 L$_{\sun}$)
have ratios between 1 and 1.5. The non-PBRS and higher luminosity PBRS tend to have higher ratios, except for the
sources that are the most evolved (i.e., HOPS 223 and HOPS 59). We therefore suggest that the visibility amplitude
ratios enable us to divide the PBRS sample into two groups. The PBRS with the
smallest visibility amplitude ratios (flattest profiles) are the youngest of the PBRS and the sources with larger
ratios (rapidly declining amplitudes) are likely more evolved, though still Class 0 protostars.

The youngest sources would then have the most compact, dense envelopes initially 
that may then be accreted rapidly due to the short
free-fall time. The flat visibility amplitudes indicate that there is likely
large amounts of mass within only a few thousand AU, implying high average densities.
A source with 2 $M_{\sun}$ of envelope material
a radius of 1500 AU has an average density of 3$\times$ 10$^7$ cm$^{-3}$,
corresponding to a local free-fall time of $\sim$10 kyr. Thus, a substantial amount
of the final protostellar mass could be accumulated in this short period of time, much less
than the expected lifetime of the Class 0 phase ($\sim$150 kyr) \citep{dunham2014}. 
Therefore, the free-fall times suggest that the Class 0 phase may begin with a
short period of rapid infall that may only last $\sim$10\% of the Class 0 phase.
Based on the number of detected sources, ST13 suggested that if the PBRS represented a 
phase of protostellar evolution distinct from the Class 0 phase, it may  only
last $\sim$ 25 kyr. Rather than necessarily being a distinct phase, we believe that the
PBRS with flat visibility amplitudes (093005, 090003, 091015, 091016, 082005, and 097002) 
are among the youngest Class 0 protostars,
and the large amount of mass on small spatial scales could indicate that they
are in a brief period of high-infall/accretion.

The PBRS that do not have flat visibility amplitudes are still young, but may be more
comparable to typical Class 0 sources. We suggest that the sources with bright 
2.9 mm flux densities, but rapidly declining visibility
amplitudes (302002, 082012, HOPS 373) are slightly more evolved that the PBRS with flat
visibility amplitudes. At least a fraction of their 
inner envelopes is likely to have been accreted onto the disk and/or protostar. The remaining sources with 
declining visibility amplitudes and low flux densities (119012, 061012, HOPS 372, 135003, 019003)
are still consistent with being young Class 0 sources.
Their cold \tbol\ values and extremely red 24 \micron\ to 70 \micron, however,
colors could result from high density envelopes, but with less overall mass. Alternatively,
they could be edge-on sources; we will further explore the properties of these sources in relation to
their outflows in an upcoming paper (Tobin et al. in prep.).

If the proposed scenario is true, then the Class 0 phase might be a two-phase process, with a
short, rapid accretion phase \citep[like a Bonnor-Ebert collapse; see ][]{foster1993}), lasting $\sim$10 - 25 kyr. This
phase is then followed by a period of slower mass assembly, for the remainder of the Class 0
phase ($\sim$100 kyr - 150 kyr), assuming a Class 0 lifetime of $\sim$ 160 kyr \citep{dunham2014}. 
This idea is consistent with the models of \citet{offner2014} that show most protostellar mass 
being accreted during the Class 0 phase, before the outflow destroys the envelope.

\subsection{Comparison to VeLLOs and candidate FSHCs}
The infrared and millimeter properties of the PBRS 
distinguish them from typical Class 0 protostars and indicate
that at least some of the PBRS may be very young Class 0 objects in a period of
high infall. Two other sub-classes of protostars identified by \textit{Spitzer} and submm/millimeter
observations are the VeLLOs and candidate FHSCs
and it is important to distinguish the PBRS from these sources
based on their millimeter properties.
First, many of the VeLLOs and candidate FHSCs are very faint at 2.9 mm.
For instance, the brightest VeLLO/candidate FHSC at 2.9 mm is 
Per-Bolo 58 with a flux density of 
13 mJy at d $\sim$ 230 pc \citep{schnee2010, enoch2010}. 
If this source was at the distance to Orion,
it would have a flux density of only $\sim$3.9 mJy and only appear
as a $\sim$4$\sigma$ detection in our data.
This flux density is less than half that
of the faintest source in the PBRS sample (119019). None of the other VeLLOs or FHSC 
candidates would be detectable at the distance to Orion with the sensitivity of our 
CARMA observations. The PBRS 061012 may be the most similar to a VeLLO, 
having the lowest luminosity, but it
has a higher 2.9 mm flux density than other VeLLOs.

A comparison to the visibility amplitudes of the VeLLOs is less straightforward. Most
VeLLOs/candidate FHSCs are not bright enough to enable analysis of their visibility amplitude profiles.
Per-Bolo-58 is found to have rapidly declining visibility amplitudes and
The candidate FHSC L1451-MMS \citep{pineda2011}, however, has 
flat visibility amplitudes at 1.3 mm, similar to some PBRS sources.
The visibility amplitudes are 30 mJy
out to $\sim$200 k$\lambda$ or 230 AU scales. At 2.9 mm, the visibility amplitudes
of this source would be 2.5 mJy, assuming $\beta$ = 1. At the distance to Orion, however,
the visibility amplitudes would be below our detection limits at 0.8 mJy. 
While the overall emitting mass of this source is much lower than the PBRS, it does have
a similar 5 k$\lambda$ to 30 k$\lambda$ flux ratio, meaning that the envelope density profile
might be similar to the most concentrated PBRS.
However, L1451-MMS is undetected at 70 \micron\ and 100 \micron, unlike the PBRS.
In summary, the millimeter properties of the PBRS, combined with the far-infrared constraints from
ST13, distinguish the PBRS from the VeLLOs and candidate FHSCs.

\section{Summary and Conclusions}

We have presented CARMA 2.9 mm dust continuum observations toward 14 PBRS  \citep{stutz2013} in the Orion 
A and B star forming regions, twelve of these protostars were first identified by \textit{Herschel}
observations. This sample of 14 PBRS also includes 135003, a new PBRS that 
discovered by \textit{Herschel} and was not included in the \citet{stutz2013} sample due to
their stringent FWHM cut-off. The inclusion
of this source increases the total number of PBRS in Orion to 19. The PBRS classification in \citet{stutz2013} required
$[24\,\mu{\rm m}]-[70\,\mu{\rm m}]$ colors or limits (in log $\lambda$F$_{\lambda}$ space) 
in excess of 1.65.  In addition, we also report the continuum properties of 4 protostars and 
1 apparent starless/pre-stellar core within the fields observed toward the PBRS.

All 14 PBRS are detected in dust continuum emission. Twelve out of 14
have flux densities $>$ 30 mJy and three have flux densities $\ge$~90
mJy. The 8 PBRS with \lbol\ $\sim$ 1 L$_{\sun}$ exhibit
higher 2.9~mm luminosities than other known protostars with similar
\lbol\ values, and therefore have characteristics not previously
identified.  The PBRS with \lbol\ $>$ 2.7 L$_{\sun}$ have comparable
2.9~mm luminosities yet lower \lbol\ values than the the brightest
sources in the more nearby regions of Perseus and Ophiuchus. Furthermore, the 2.9~mm luminosity does not
strongly correlate with \lbol.  This lack of correlation indicates
that the 2.9~mm luminosity is not strongly dependent on either the the
central protostellar luminosity or the accretion luminosity.

Six PBRS sources (097002, 090003, 093005, 082005, 091015, and 091016)
have flat 2.9~mm visibility amplitudes (and 5~k$\lambda$ to
30~k$\lambda$ visibility amplitude ratios of less than 2).  As a
consequence, more than $\sim\,$50\% of the total flux density arises
from scales that are smaller than 7\arcsec\ ($\sim\,$3000~AU) in
diameter.  This behavior indicates either steep envelope density
profiles or the presence of significant mass contained within a
compact, unresolved structure. We suggest that these particular PBRS
are the youngest of the sample and may be in a brief period of high
infall rate. Indeed, the average density on scales $<$~3000~AU implies
local free-fall times of $\sim\,$10 kyr, in agreement with independent
life-time estimates based on the ratio of PBRS to protostars
\citep{stutz2013}.  The PBRS with large 2.9~mm flux densities but
rapidly declining visibility amplitudes (302002, 082012, HOPS 372) are
still considered to be young Class~0 protostars, but may be more
evolved than the PBRS with flat visibility amplitudes. The sources
with lower 2.9~mm flux densities and declining visibility amplitudes
(119012, 061012, HOPS 372, 135003, 019003) are also still consistent
with being Class~0 sources, but may have edge-on orientations and/or
lower envelope masses.

To better characterize the density profiles of the sample, we compare
the observed visibility amplitudes of the sources to Hyperion
radiative transfer models of axisymmetric envelopes with varying
radial density profiles and unresolved components (represented by a
disk component).  We also compare with rotating collapse models with
various centrifugal radii and disk masses.  We find that without an
unresolved component to the emission, the flat visibility amplitude
sources are most consistent with a $\rho$ $\propto$ R$^{-2.5}$ radial
density profile.  If a compact structure is massive enough, however,
then all three envelope density profiles tested here ($\rho$ $\propto$ 
R$^{-1.5,-2.0,-2.5}$) are able to provide a
reasonable match to the data.  Thus, with the current data we cannot
distinguish between these two scenarios and higher resolution data are
required to understand the nature of these sources. Furthermore,
sources with more rapidly decreasing visibility amplitudes may be
consistent with shallower density profiles and are inconsistent with
having a massive unresolved component.

While the PBRS occupy a narrow range to \tbol, their 2.9~mm flux
densities and visibility amplitude profiles show a large amount of
heterogeneity suggesting that they are not all in exactly the same
evolutionary stage.  We suggest an evolutionary trend in which the
sources with flat visibility amplitude profiles are the youngest and
perhaps have a dense inner envelope that may be rapidly accreted. The
sources large 2.9~mm flux densities and rapidly declining visibility
amplitudes may be slightly more evolved than those with flat
visibility amplitudes. Moreover, the sources with flat visibility
amplitude profiles also tend to have lower \lbol\ values than those
with rapidly declining visibility amplitudes, consistent with the
expected evolutionary trend of increasing \lbol. The PBRS also draw a
sharp contrast with candidate FHSCs and VeLLOs. The PBRS have higher
\lbol\ and larger 2.9 mm luminosities than all of the candidate FHSCs
and VeLLOs; at the distance to Orion, none of the known candidate
FHSCs or VeLLOs would have been confidently detected.

In summary, the PBRS have properties that are consistent with placing
them among the youngest Class~0 protostars.  Their millimeter and
infrared properties distinguish them from typical Class~0 protostars,
as well as from candidate FHSCs and VeLLOs. While the data presented
here have enabled us to postulate a tentative evolutionary scenario,
further characterization at higher resolutions and in molecular lines
will be necessary to more firmly establish their place in the context
of the star formation process.

The authors wish to thank the anonymous referee for constructive comments
that improved quality and clarity of the manuscript.
J.T. acknowledges support provided by NASA through Hubble Fellowship 
grant \#HST-HF-51300.01-A awarded by the Space Telescope Science Institute, which is 
operated by the Association of Universities for Research in Astronomy, 
Inc., for NASA, under contract NAS 5-26555. J.T. also acknowledges funding from
\textit{Herschel} OT2 JPL grant \#1458263. 
The work of A.M.S. and S.E.R. was supported by the Deutsche
Forschungsgemeinschaft priority program 1573 ('Physics of the
Interstellar Medium'). Support for STM and WJF was provided by 
NASA through awards issued by JPL/Caltech. Support for CARMA construction 
was derived from the states of Illinois, California, and Maryland, the 
James S. McDonnell Foundation, the Gordon and Betty Moore Foundation, the Kenneth 
T. and Eileen L. Norris Foundation, the University of Chicago, the Associates of the 
California Institute of Technology, and the National Science Foundation. Ongoing 
CARMA development and operations are supported by the National Science Foundation 
under a cooperative agreement, and by the CARMA partner universities. 
The National Radio Astronomy 
Observatory is a facility of the National Science Foundation 
operated under cooperative agreement by Associated Universities, Inc.

\appendix
\section{Notes on Individual Sources}
\subsection{135003}

The PBRS source 135003 is located in the OMC2/3 region of the Orion A cloud.
 This source was not originally included in the PBRS sample of ST13, due to 
it having a 70 \micron\ full-width at half maximum size slightly greater than
7\farcs8. A multi-wavelength plot of this source and its SED are given Figures \ref{135003} and \ref{sed-135003}.
The 24 \micron\ to 70 \micron\ color 
(log($\lambda$F$_{\lambda}$(70\micron)/$\lambda$F$_{\lambda}$(24\micron)) for 135003 is 2.44 and
it has \lbol\ = 11.2 L$_{\sun}$, \tbol\ = 34.6 K, L$_{submm}$/\lbol\ = 0.175, and
an SED peak at $\sim$121 \micron. These values are derived from modified blackbody 
fitting, the same method used in ST13. The photometry for 135003 is given in Table 4.
The brighter, neighboring source is HOPS 59, a Class I 
protostar and separated by 18\arcsec\ ($\sim$7600 AU); HOPS 59 is bright
both at 2.9 mm and in the infrared. There
is another 3$\sigma$ source located between 135003 and HOPS 59, 
which may be a real source given that there is an IRAC
point source detected at this position at 3.6 \micron\ and 4.5 \micron. The northeastern
extension of emission from 135003 is real and has been observed in
3 mm single-dish maps of the region \citep{schnee2014}, in
addition to the 350 \micron\ and 870 \micron\ maps showing similarly extended structure.
The visibility amplitudes of this source appear to flatten
toward the longer baselines, while there is some up and down
variation likely due to the extended emission and surrounding sources. 
Nevertheless, the overall flux density
at large uv-distances is less than other PBRS.

\subsection{093005}
The reddest PBRS in 24 \micron\ to 70 \micron\ color, 093005, is located at the intersection of 
three filaments and $\sim$110\arcsec\ north of its nearby neighbor of HOPS 373 (ST13).
At wavelengths shorter than 70 \micron, 093005 was only 
detected in \textit{Spitzer} 4.5 \micron\ imaging (ST13). 
The dust continuum emission from this source is quite compact, only $\sim$1.2$\times$
fainter at 30 k$\lambda$ versus 5 k$\lambda$ and the visibility amplitudes are very flat out to $\sim$ 80 k$\lambda$.

\subsection{HOPS 373}

HOPS 373 is the close neighbor of 093005, located 110\arcsec\ 
to the south. The dust continuum emission observed in D-configuration
only showed some asymmetry and the combined D and C configuration
data resolve a second component, separated by 4\arcsec.
The continuum emission from HOPS 373 is fainter than 093005 
and the visibility amplitudes fall quickly with increasing uv-distance, 
reaching zero at 50 k$\lambda$ and then rise again, showing the expected 
pattern for two sources separated by 4\arcsec.

\subsection{090003}

The PBRS 090003 (also called Orion B9 SMM 3; \citet{miettinen2012a}) is 
located in a loose filamentary complex with several protostars and 
starless cores over a 0.5 pc region \citep{miettinen2012a}.
Much like 093005, its only detection shortward of 24 \micron\ is at 4.5 \micron, where there
is a small feature offset from the location of the protostar, 
indicative of shocked H$_2$ emission \citep{miettinen2012a,stutz2013}. 

The envelope as observed at 870 \micron\ is extended to the east
and at 350 \micron\ there are a 
few sub-peaks associated with the extended 870 \micron\
emission \citep{miettinen2012a}. We do not, however, detect 2.9 mm continuum emission from these sub-peaks,
indicating that they do not harbor compact protostellar sources. The continuum
emission from 090003 has a very similar visibility amplitude 
profile to 093005, indicating that this source
is also dominated by small-scale emission. The source 090003 is the 
second brightest in the sample at 2.9 mm.

\subsection{082012 and HOPS 372}

The brightest continuum source in the sample is 082012; 
the 2.9 mm continuum luminosity of 082012 is about the same as that of the
well-studied multiple protostellar system IRAS 16293-2422. Similarly, 082012
has visibility amplitudes that are falling quickly as a function of uv-distance, 
indicating that the source is not dominated by compact emission.
Moreover, 082012 (\lbol\ = 6.3 L$_{\sun}$) has 
a lower luminosity than other protostars with similar 
envelope characteristics. Only $\sim$20\arcsec\ away from 082012 is HOPS 372, another PBRS source, but with much lower 
overall continuum flux.

\subsection{019003}

The PBRS source 019003 is also located in the OMC 2/3 region, northward of 135003. 
We detect four sources at 2.9 mm surrounding 019003 (Figure 1):
HOPS 71 farthest to the east, HOPS 68 to the south, and 019003 itself, which resolves into two sources 
denoted 019003 A and 019003 B.
The source 019003 A is most closely associated with the 70 \micron\ source detected by ST13. 
The 70 \micron\ source, however, is offset by $\sim$ 3\arcsec\ to the west, as are the 
4.5 \micron\ and 24 \micron\ sources. On the other hand, 019003 B appears to be starless, with no detections at any wavelength shorter that 160 \micron.

\subsection{HOPS 68}

The protostar HOPS 68 is detected at the edge of the 019003 field. This source was previously recognized
for the appearance of crystalline silicates in its \textit{Spitzer} IRS spectrum \citep{poteet2011}.

\subsection{302002}

The source 302002 is located at the end of an isolated filamentary structure in NGC 2068,
$\sim$20\arcsec\ to the south of a Class I source (HOPS 331).
 The 2.9 mm continuum is well-detected toward 302002, whereas HOPS 331 is 
undetected. The visibility amplitudes toward this source are falling with 
increasing uv-distance, similar to those of
082012.

\subsection{061012 and HOPS 223}

The source 061012 is located very near the outbursting protostar 
V2775 Ori (HOPS 223; also detected at 2.9 mm) in the L1641 region \citep{fischer2012}.
This source is one of the faintest PBRS in the sample at 17 mJy; indeed, only 119019 is slightly fainter. 
The visibility amplitudes appear rather flat, but become dominated
by noise at uv-distances larger than 20 k$\lambda$. Hence, its nature is uncertain.

\subsection{091015 and 091016}

The two sources 091015 and 091016 are close neighbors in NGC 2068, 
the former being $\sim$40\arcsec\ west of the latter.
These sources are completely undetected at wavelengths shortward of 70 \micron.
The visibility amplitudes are flat across 
all observed uv-distances toward both sources in Figure \ref{uvamps}, with 
091015 declining a bit more rapidly than 091016.

\subsection{082005}

The source 082005 is located about 4\arcmin\ south of 082012, and these sources are connected by an 
apparent filamentary structure seen at 160 \micron\ and 870 \micron. This source 
is also undetected at wavelengths shorter than 70 \micron, like 091015 and 091016. 
The visibility amplitude data for the
dust continuum are rather flat like some of the other deeply embedded sources, but still decline
more rapidly than the flattest sources.

\subsection{097002}

The source 097002 is found near a bright 4.5 \micron\ and 24 \micron\ source seen in \textit{Spitzer}\
data, as shown by ST13. At 70 \micron, the neighboring 24 \micron\ source 
is undetected, but 097002 is quite apparent.
We detect 097002 as a bright 2.9 mm continuum source with very flat visibility amplitudes, but 
we do not detect the neighboring source. 
Much like 090003 and 093005, the visibility amplitudes toward this source are very flat.

\subsection{119019}

The source 119019 is the faintest PBRS source in terms of 2.9 mm continuum 
emission and appears more similar to the protostars observed in
the \citet{tobin2011} sample rather than the PBRS sources.

\begin{small}
\bibliographystyle{apj}
\bibliography{ms}
\end{small}

\begin{figure}
\begin{center}
\includegraphics[scale=0.8]{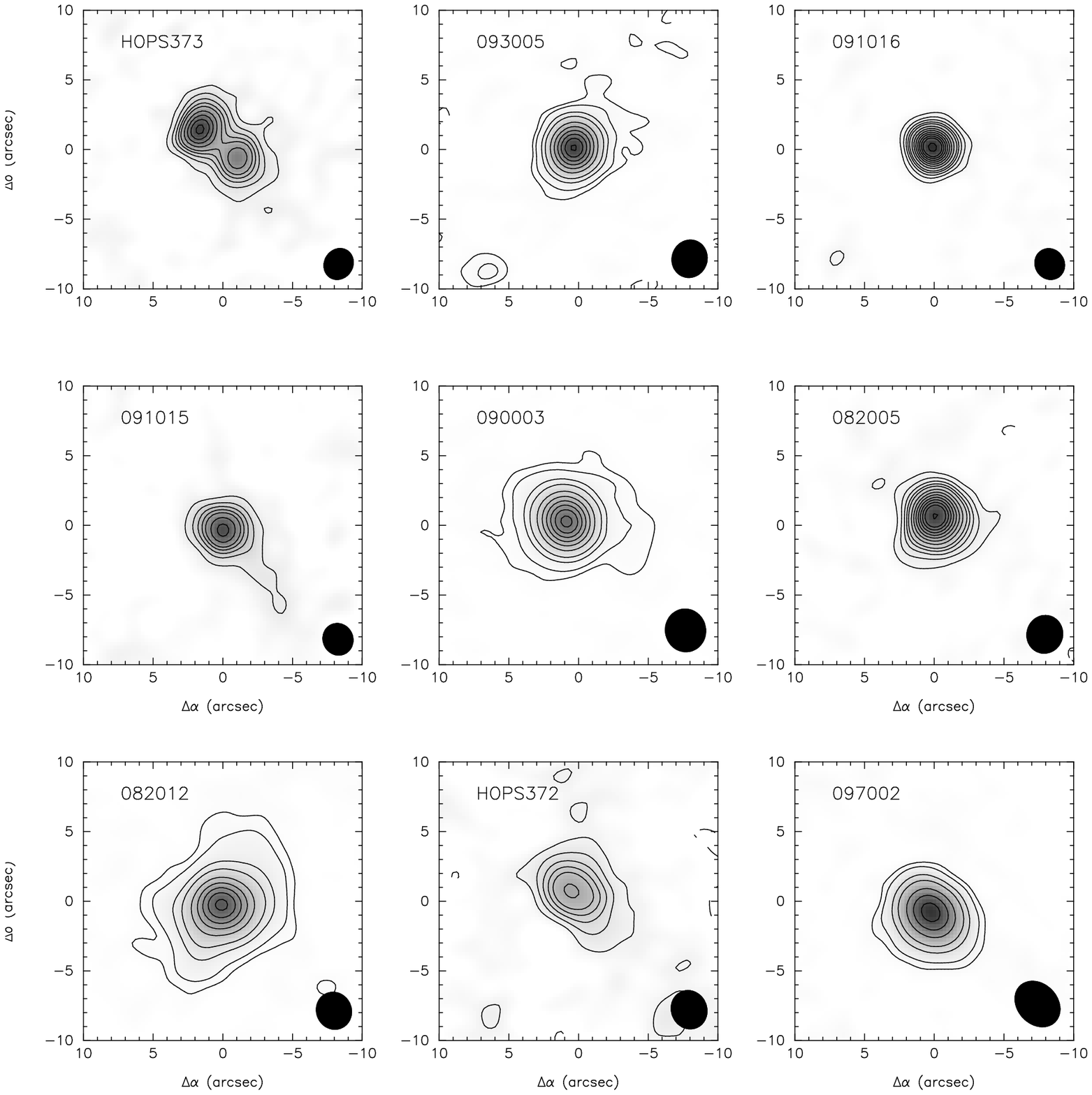}

\end{center}
\caption{Continuum images at 2.9 mm for the observed PBRS sample and sources within the field of view.
 While many appear mostly round, some have marginally resolved  emission 
and Table 3 shows that the peak flux densities are less than
the integrated flux densities. The sources in Figure 1a are a combination of D and C configuration data and the
sources with only D-configuration data are shown in Figure 1b. 
The contours are [-3, 3, 6, 9, 12, 15, 20, 25, 30, 35, 40, 45, 50, 60, ...] $\times$ $\sigma$
for all sources except 093005, 090003, and 082012 where the contours are [-3, 3, 6, 15, 25, 45, 65, 85, ...] $\times$ $\sigma$;
$\sigma$ is given for each source in Table 2.}

\end{figure}
\label{continuum}

\begin{figure}
\figurenum{1b}
\begin{center}

\includegraphics[scale=0.8]{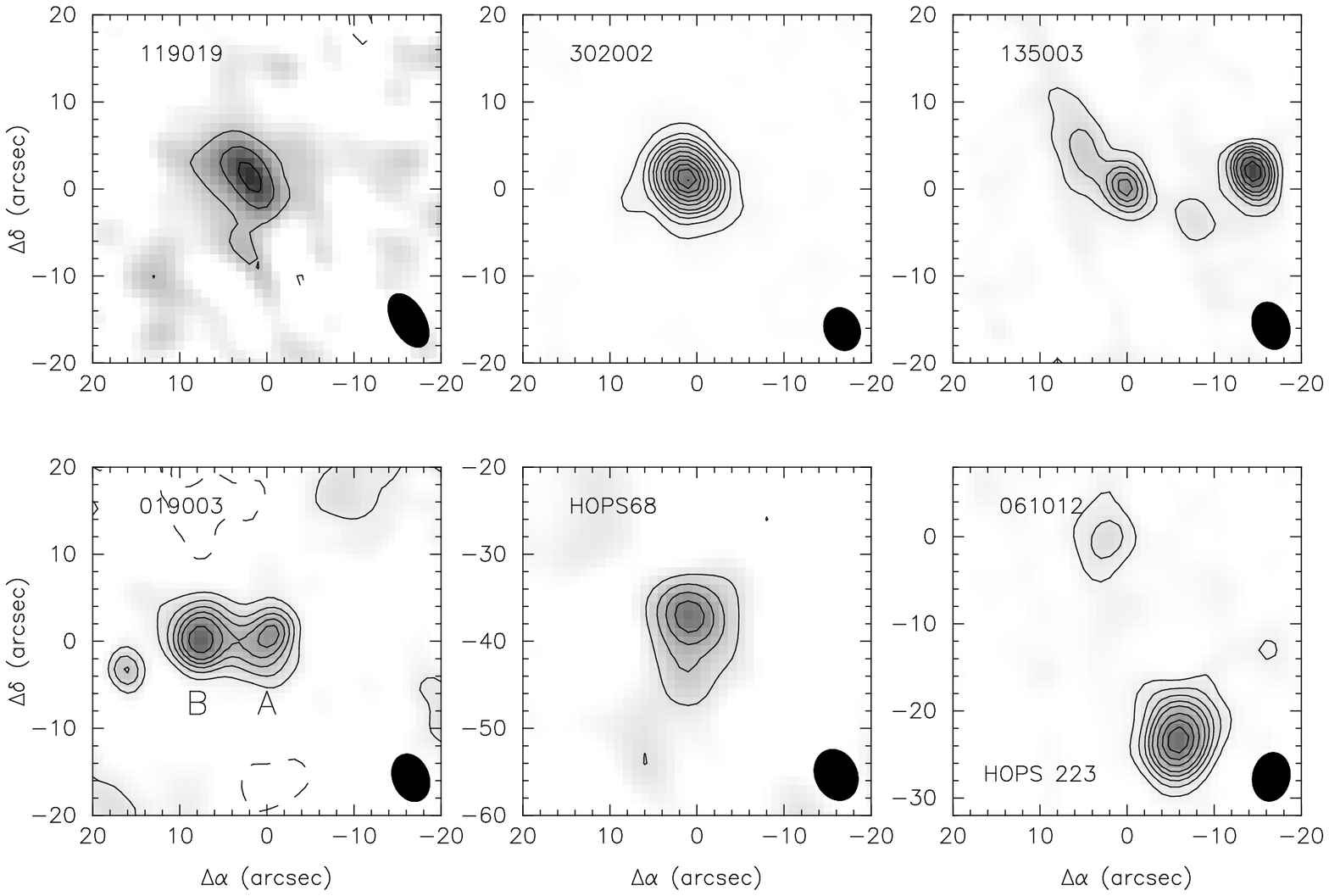}

\end{center}
\caption{}
\end{figure}

\begin{figure}
\begin{center}
\includegraphics[scale=0.35]{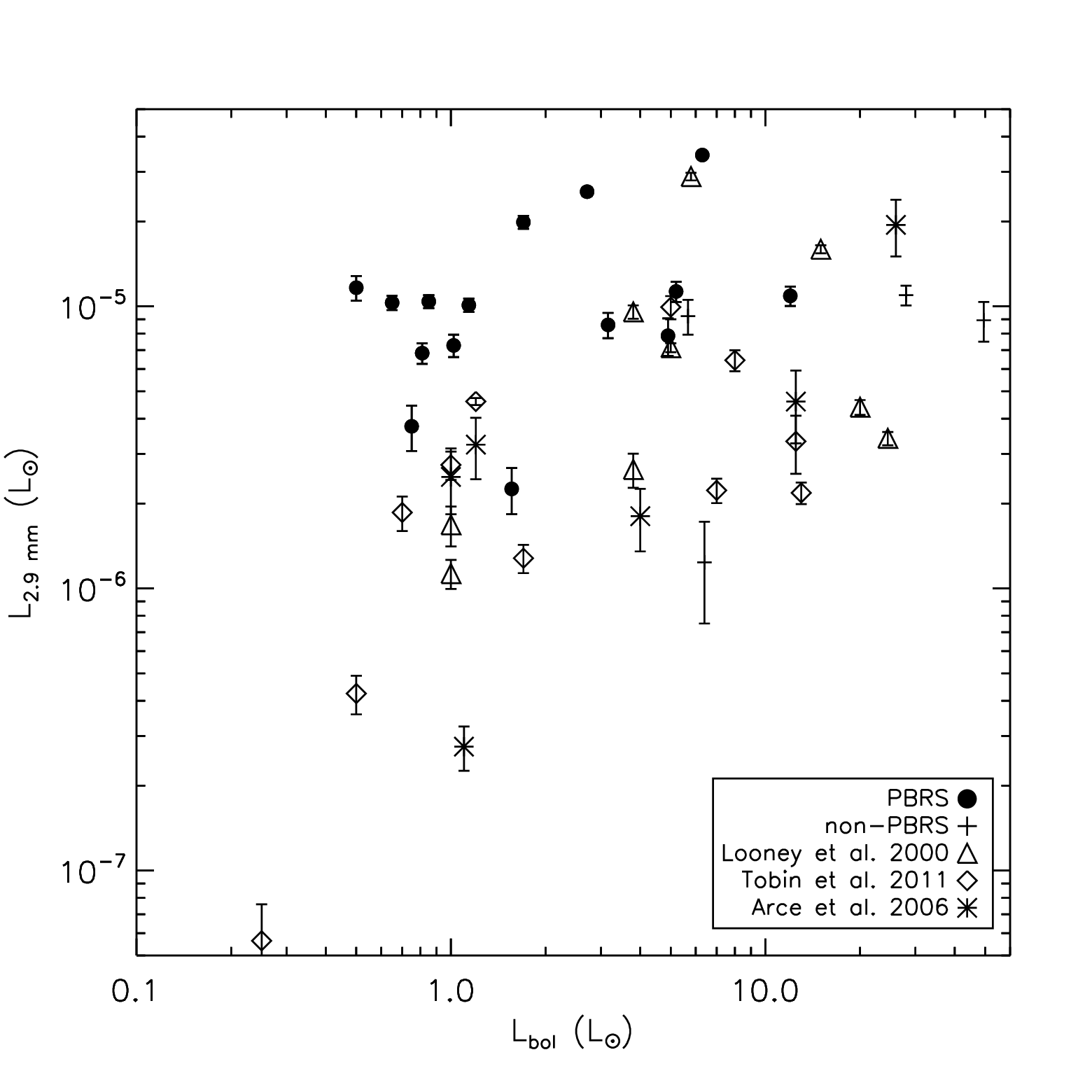}
\includegraphics[scale=0.35]{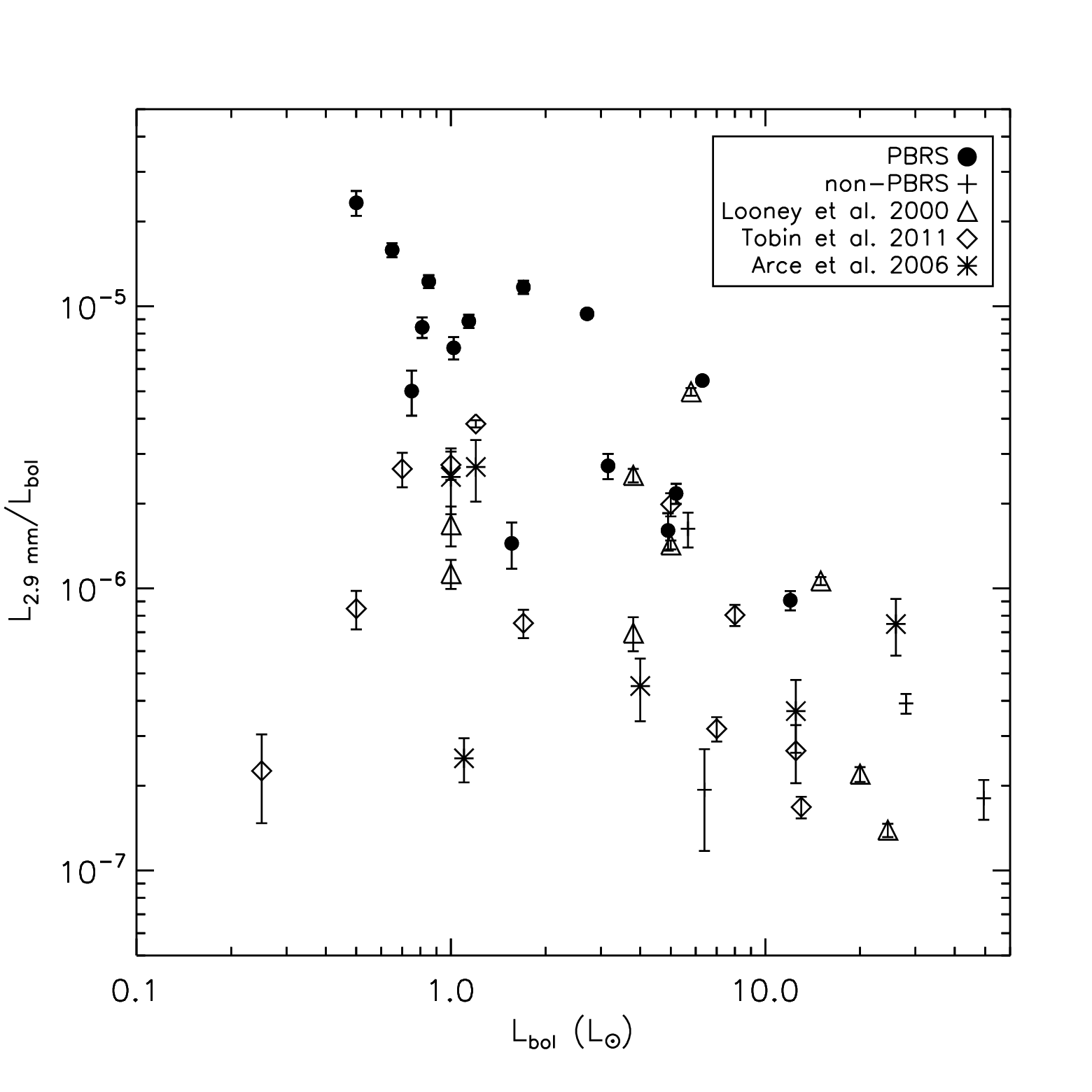}
\end{center}
\caption{Plots of the 2.9 mm luminosity (L$_{2.9mm}$) versus \lbol\ (top) 
and the L$_{2.9mm}$/\lbol\ ratio versus \lbol\ (bottom). 
We also include sources observed by \citet{tobin2011} at 3.4 mm,
by \citet{looney2000} at 2.7 mm, and by \citet{arce2006} at 2.7 mm.
For luminosities of about 1 $L_{\sun}$, the PBRS have the highest 
2.9 mm luminosity ratios and at higher luminosities
they are comparable to or larger than those from \citet{looney2000}.
Note that we have rescaled the flux densities from
the literature to account for the spectral slopes, assuming $\beta$ = 1, 
the 3.4 mm flux densities are increased by a factor
of 1.6 and the 2.7 mm flux densities are decreased by a factor of 0.8. Note that the
scaling does not affect the result of the PBRS having among the highest L$_{2.9mm}$ values for
a given \lbol.
}
\label{mmflux}
\end{figure}

\begin{figure}
\begin{center}
\includegraphics[scale=0.6]{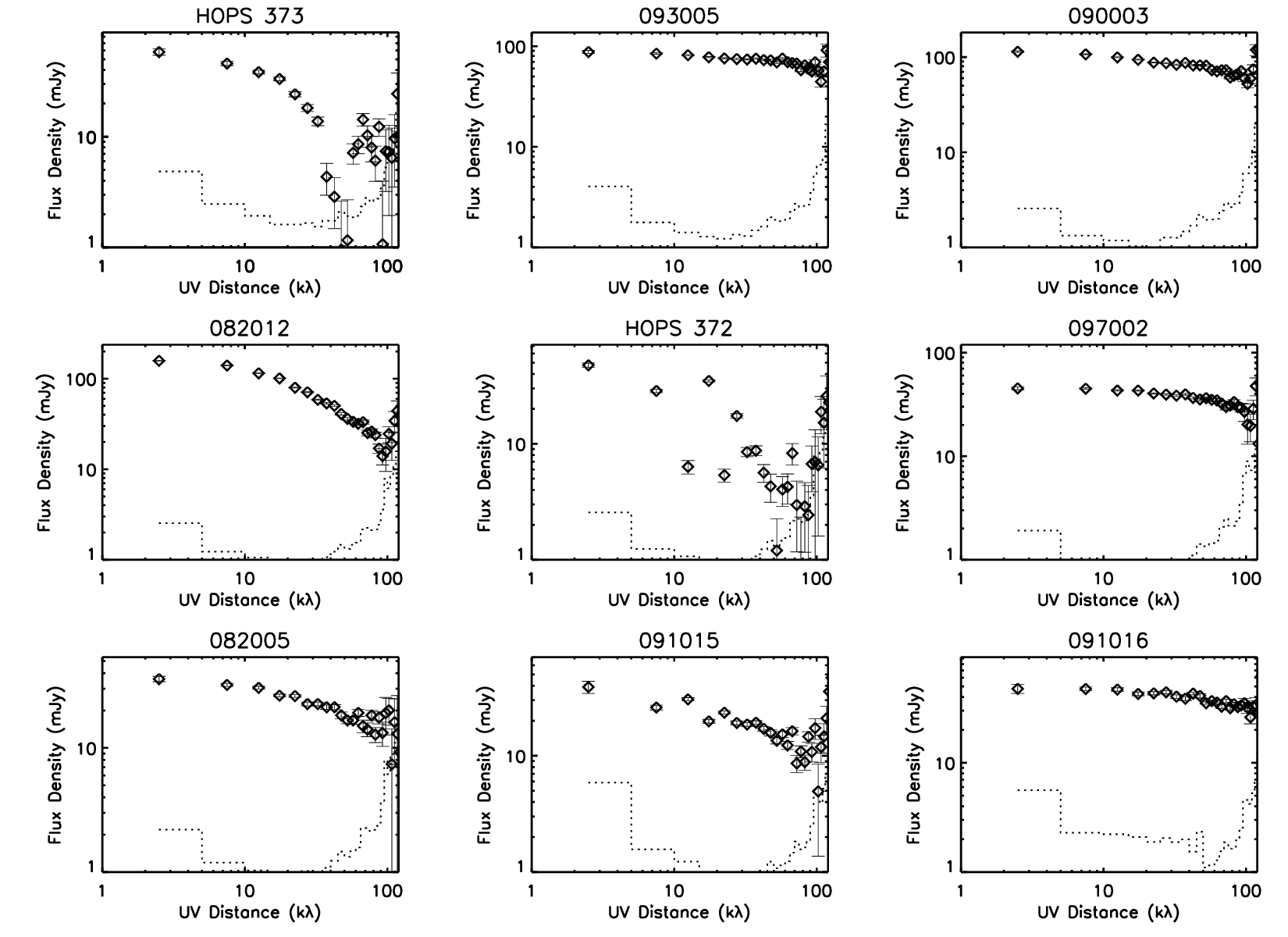}

\end{center}
\caption{Visibility amplitude versus uv-distance plots for all the sources. Flatter visibility amplitudes with increasing projected uv-distance
indicate that the flux density is dominated by compact, unresolved emission (e.g., 093005, 090003, 082005, 091015, 091016). Visibility
amplitudes that are decreasing rapidly with increased uv-distance indicate that there is more flux on larger spatial scales, relative
to a compact unresolved component (e.g., 082012, 302002, HOPS 373, HOPS 223). The light dashed line in each plot is the expected
visibility amplitude that would be measured from noise alone. Sources only observed in 
D-configuration with shorter uv-distances are shown in Figure 3b. }
\label{uvamps}
\end{figure}

\begin{figure}
\figurenum{3b}
\begin{center}

\includegraphics[scale=0.6]{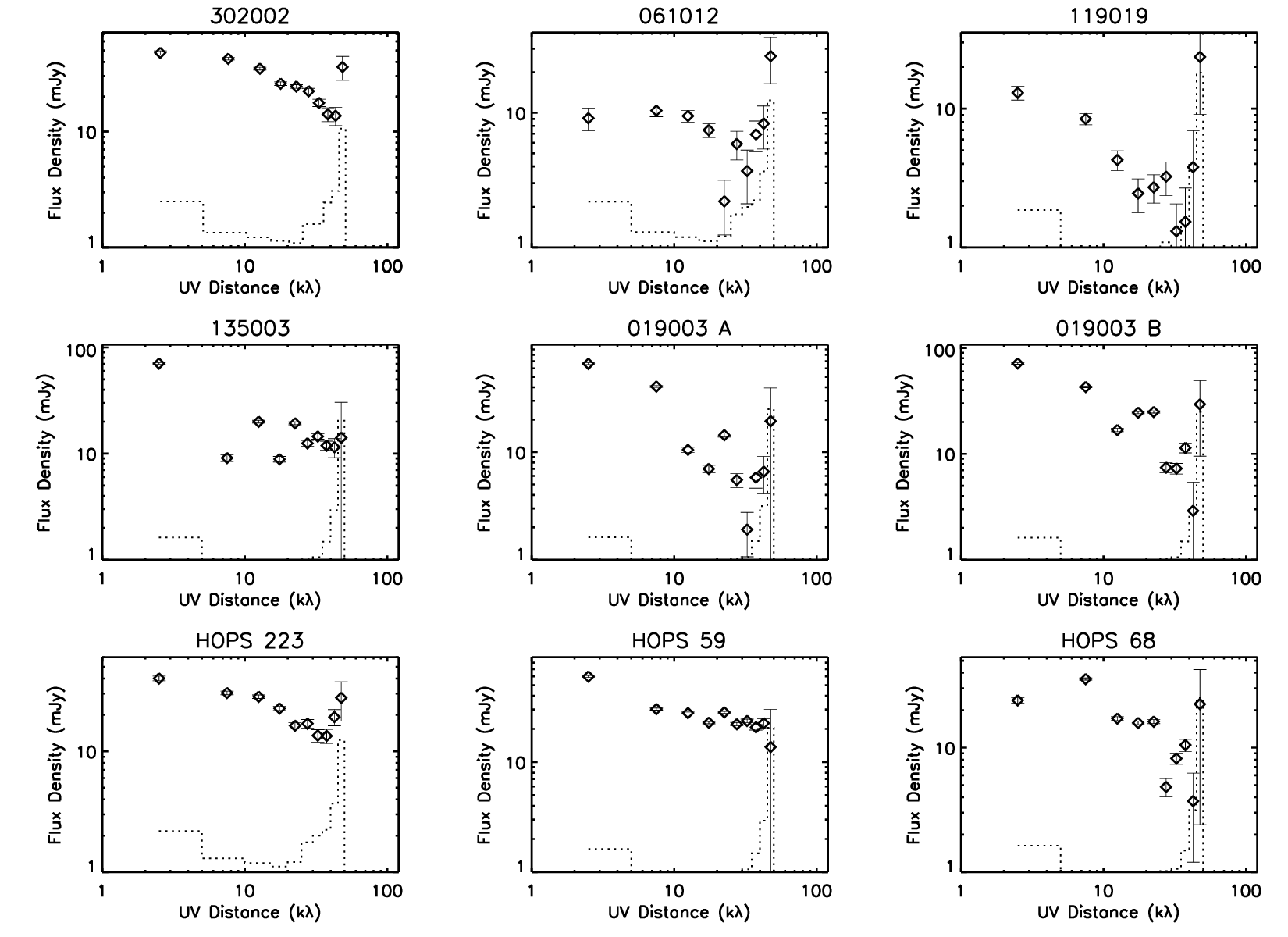}

\end{center}
\caption{}

\end{figure}

\begin{figure}
\begin{center}
\includegraphics[scale=0.4]{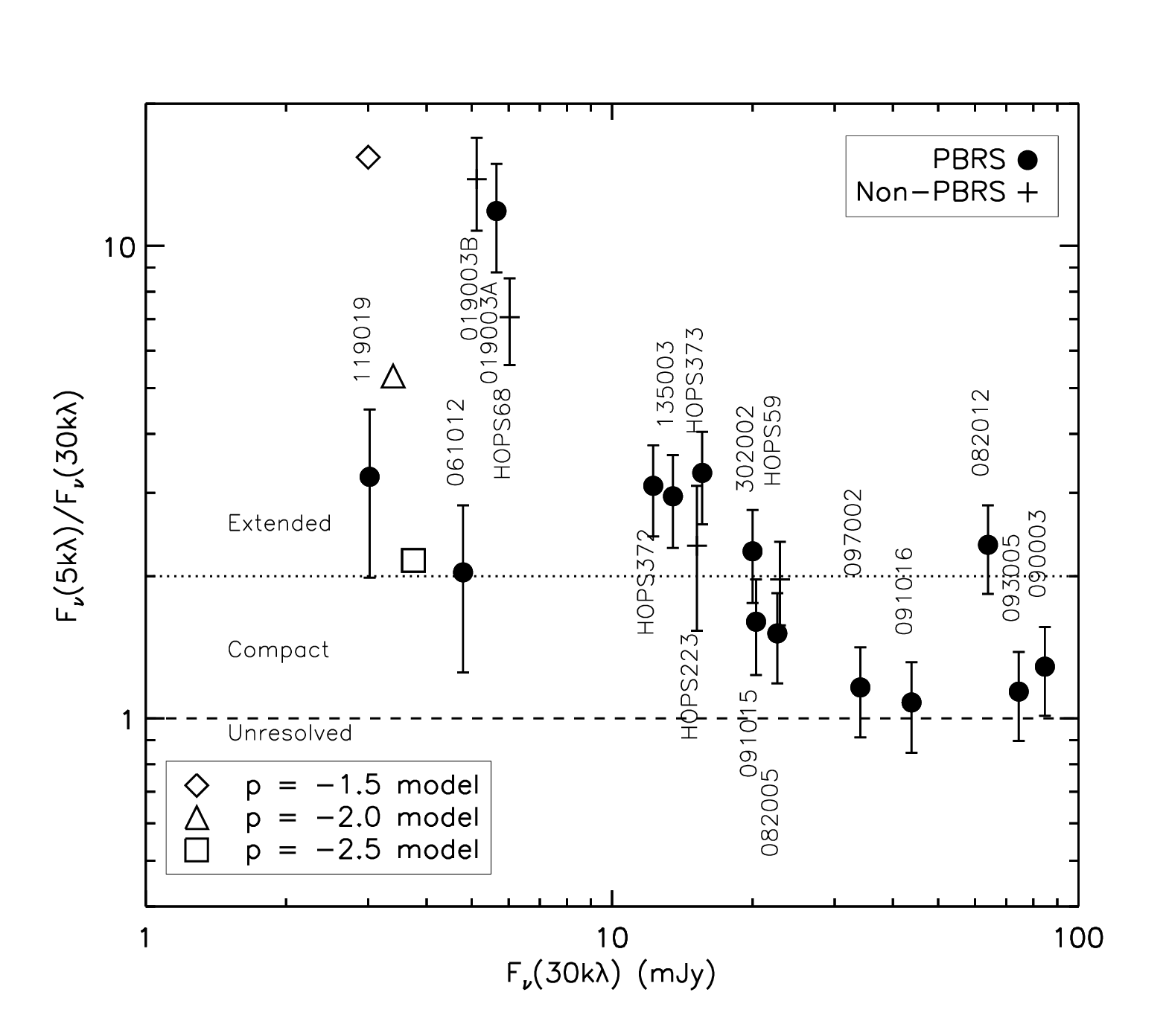}
\includegraphics[scale=0.4]{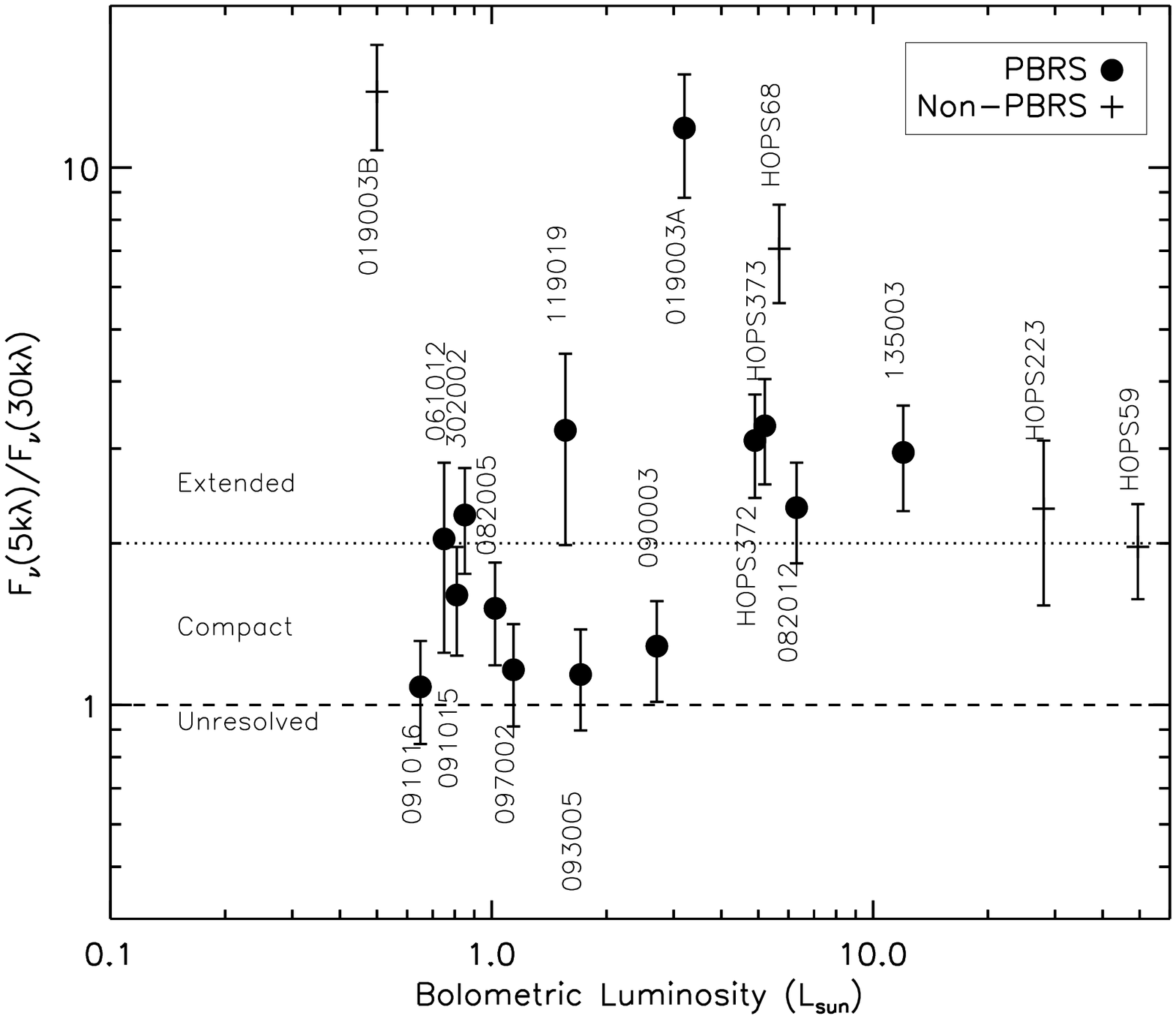}
\end{center}
\caption{Plot of the flux density ratio at 5 k$\lambda$ and 30 k$\lambda$  to the flux density at 30 k$\lambda$
(top) and a plot of $L_{bol}$ versus the flux density ratio at 
5 k$\lambda$ and 30 k$\lambda$ versus $L_{bol}$ (bottom). The higher the ratio,
the more flux is resolved out at the larger uv-distances.
This plot enables us to separate the PBRS and other protostars into two groups, one
with low flux ratios, meaning that most emission is from small spatial scales and the other where most
emission is from larger spatial scales. There is a trend for the lower luminosity PBRS
to have lower 5 k$\lambda$ to 30 k$\lambda$ flux ratios, while more luminous sources
have higher ratios. The fine dotted line is for a 5 k$\lambda$ to 30 k$\lambda$ flux ratio of 2, about
the limit where sources significantly deviate from a flat appearance in Figure 3 and the dashed line simply
is a flux density ratio of unity for reference. Ratios from model envelopes with density profiles proportional
to R$^{-p}$ where p = -1.5, -2.0, and -2.5 are shown in the top panel as a diamond, triangle, and square, respectively. 
}
\label{uvratio}
\end{figure}

\begin{figure}
\begin{center}
\includegraphics[scale=0.27]{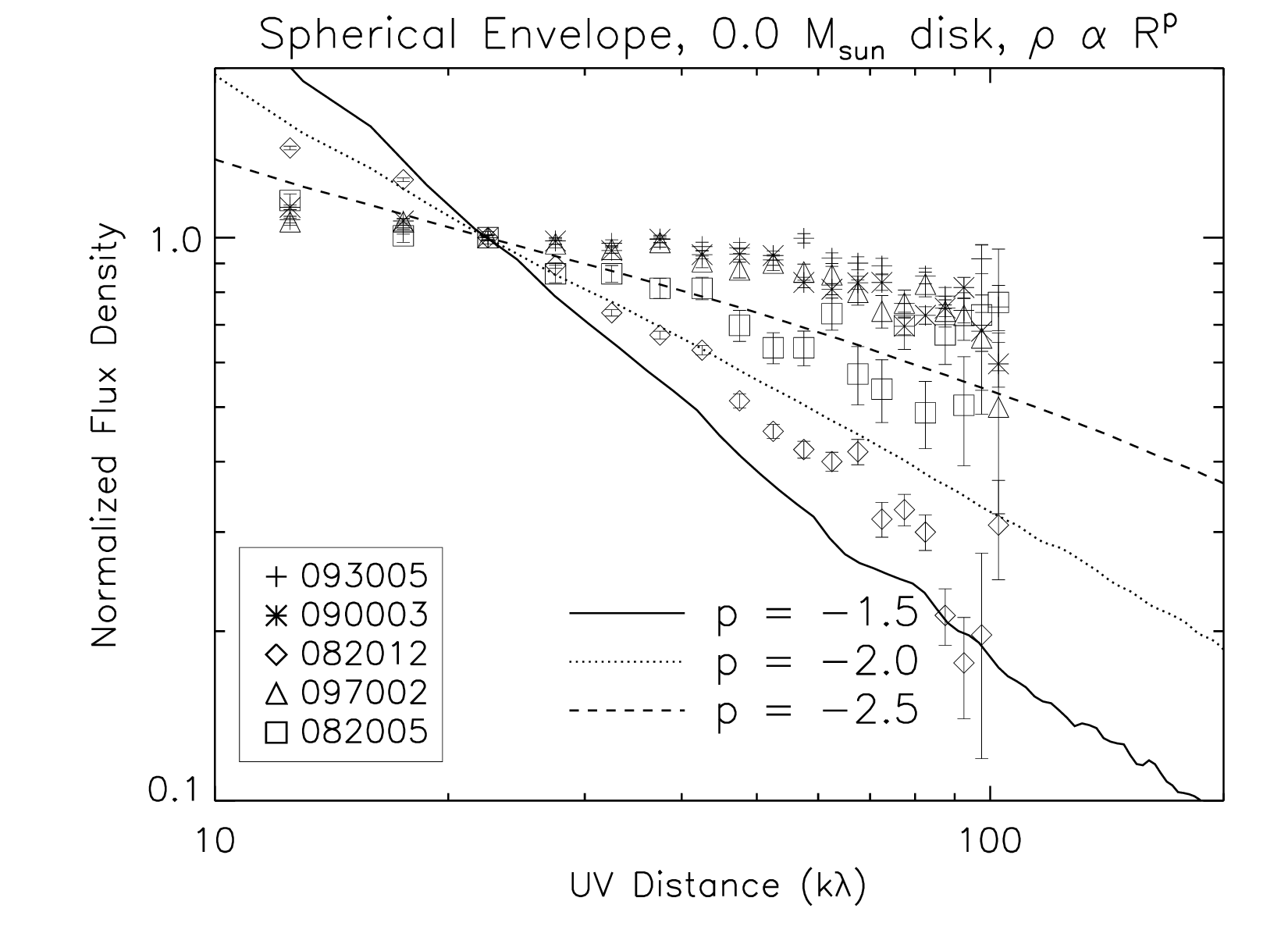}
\includegraphics[scale=0.27]{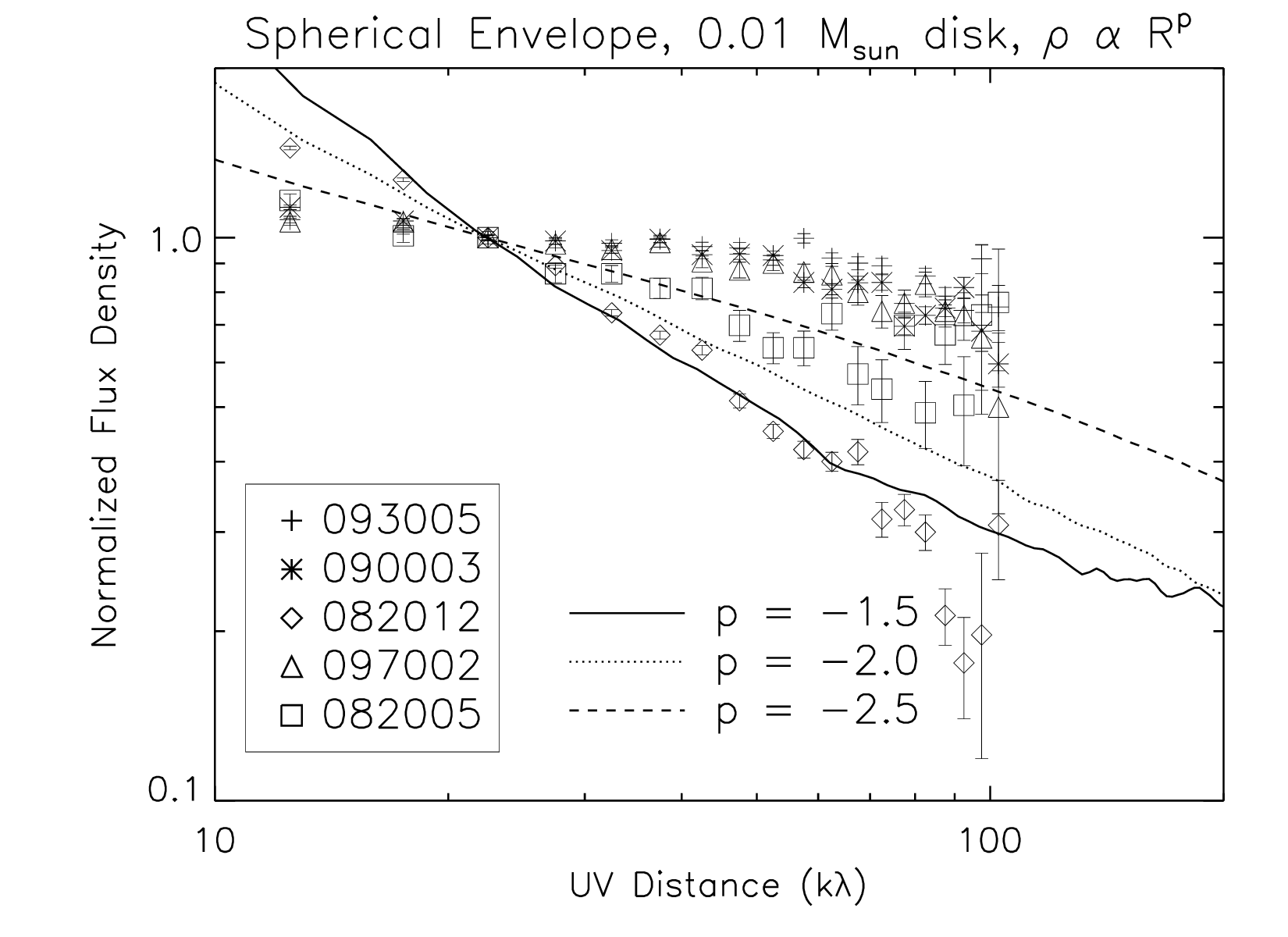}
\includegraphics[scale=0.27]{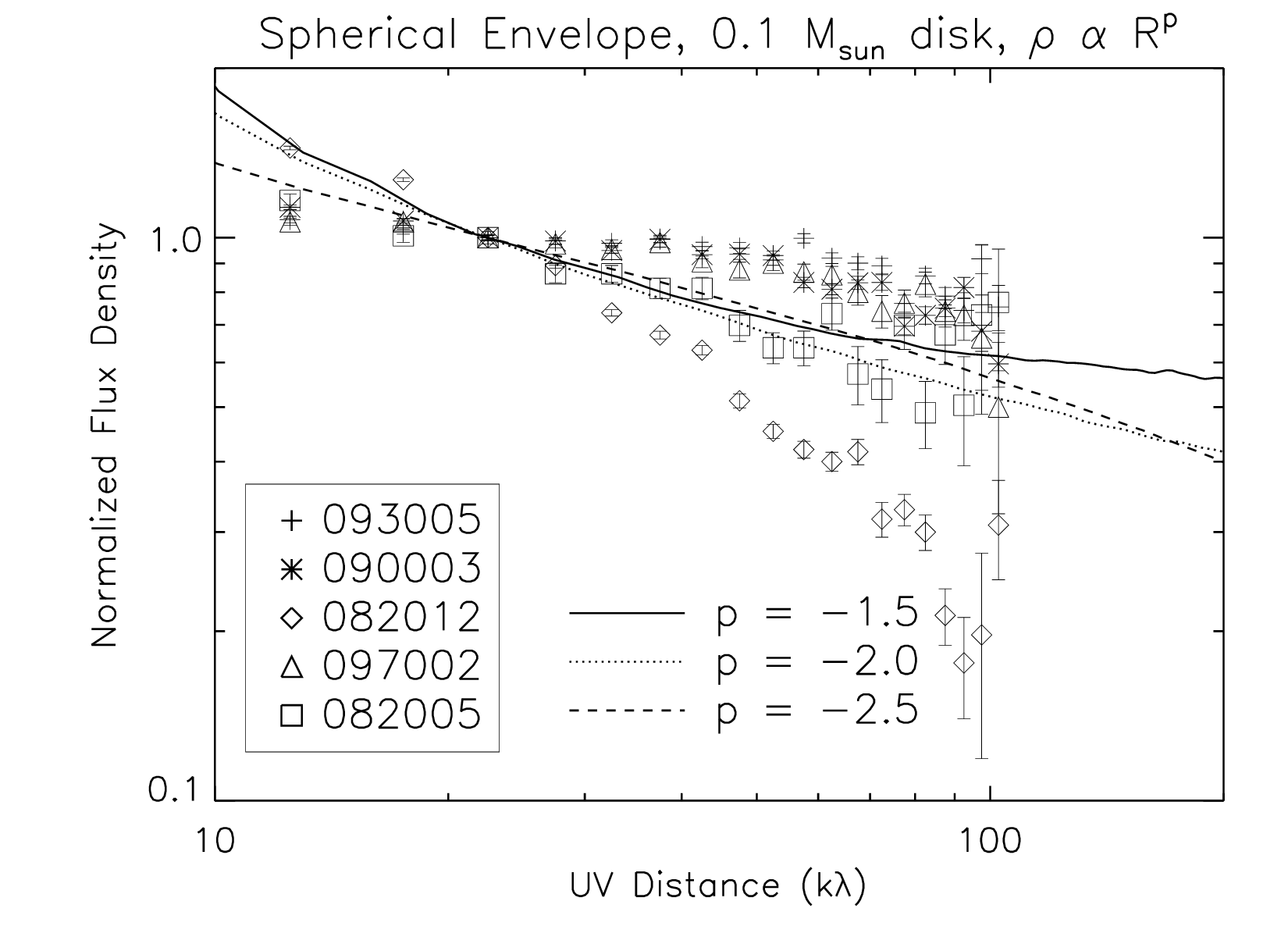}
\end{center}
\caption{Plots of normalized visibility amplitudes versus uv-distance for models 
and a few representative sources overlaid. We 
plot radial density profiles ($\rho$ $\propto$ $R^{-1.5, -2.0, -2.5}$) and 
compact object (disk) masses of 0.0 $M_{\sun}$, 0.01 $M_{\sun}$, and 0.1 $M_{\sun}$
in each plot. The flux densities are normalized at 22.5 k$\lambda$ (9.2\arcsec; 3850 AU) 
to limit the contribution of flux external to the envelopes and only probe
internal structure. With no compact structure (i.e., no disk), the
 sources with flat
visibility amplitudes are consistent with either having a density profile as steep as $\rho$ $\propto$ $R^{-2.5}$ or
being dominated by unresolved  structure, while 082012 is inconsistent with being
dominated by unresolved structure.
 }
\label{uvcomps}
\end{figure}

\begin{figure}
\begin{center}

\includegraphics[scale=0.27]{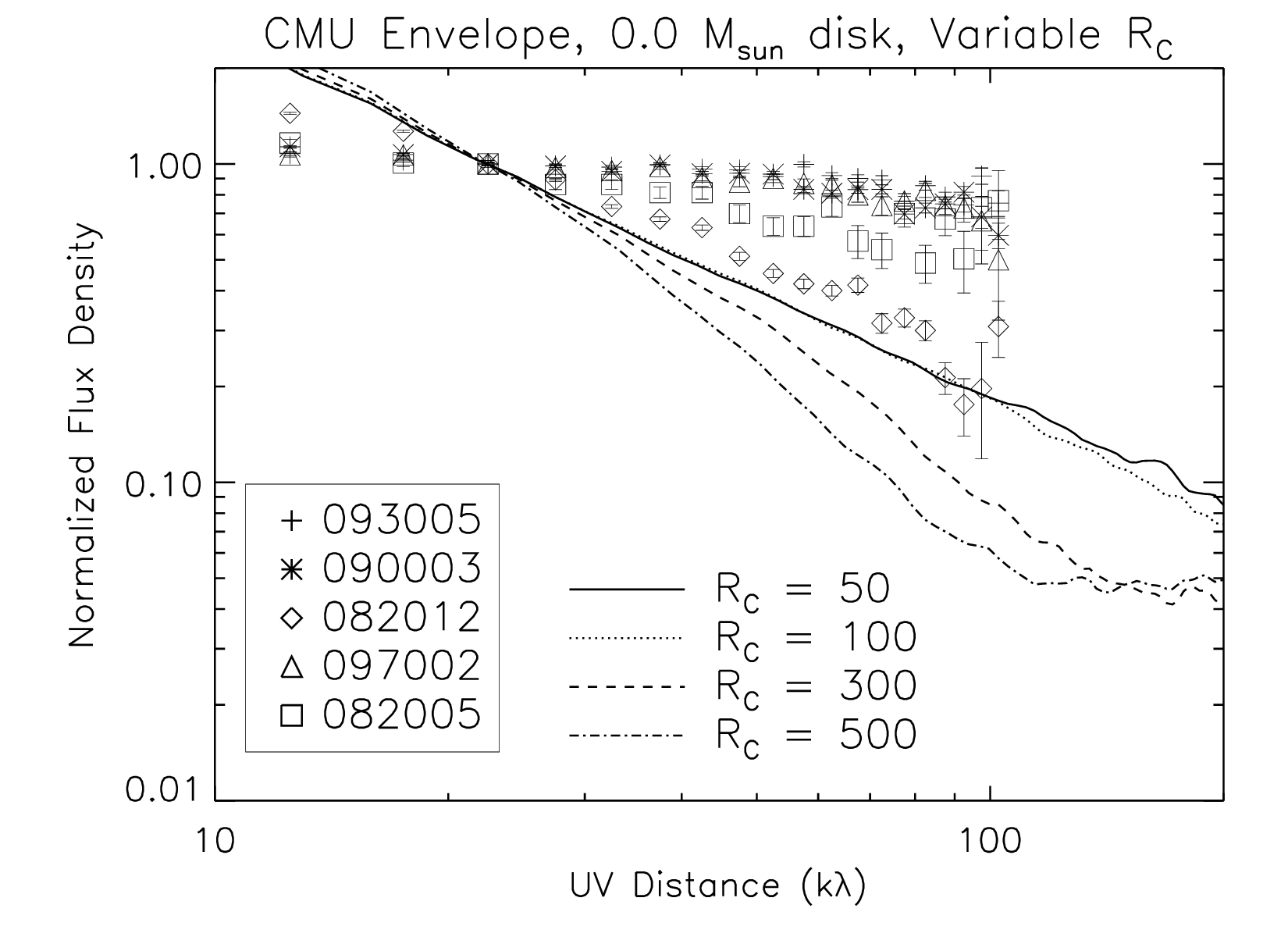}
\includegraphics[scale=0.27]{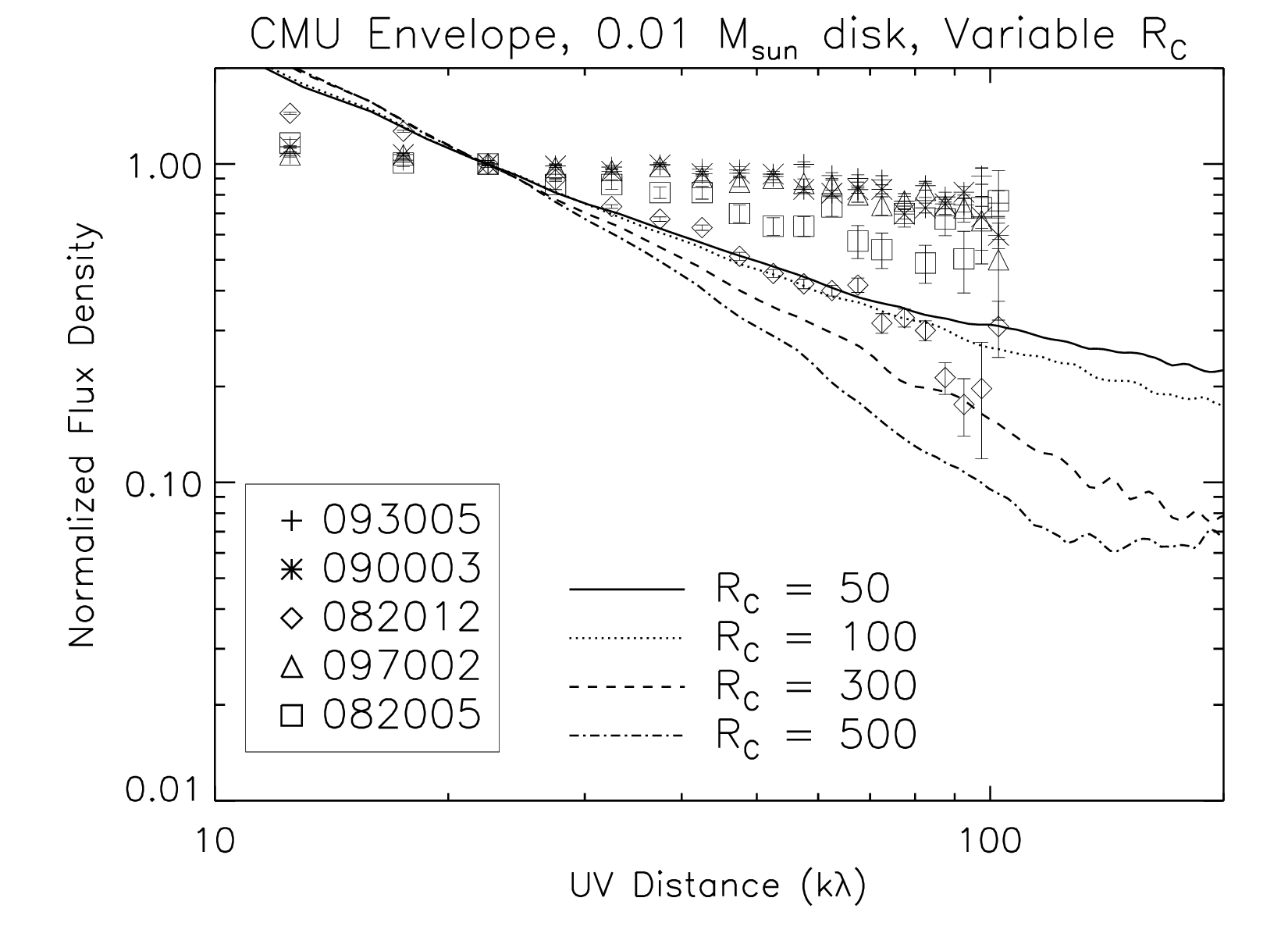}
\includegraphics[scale=0.27]{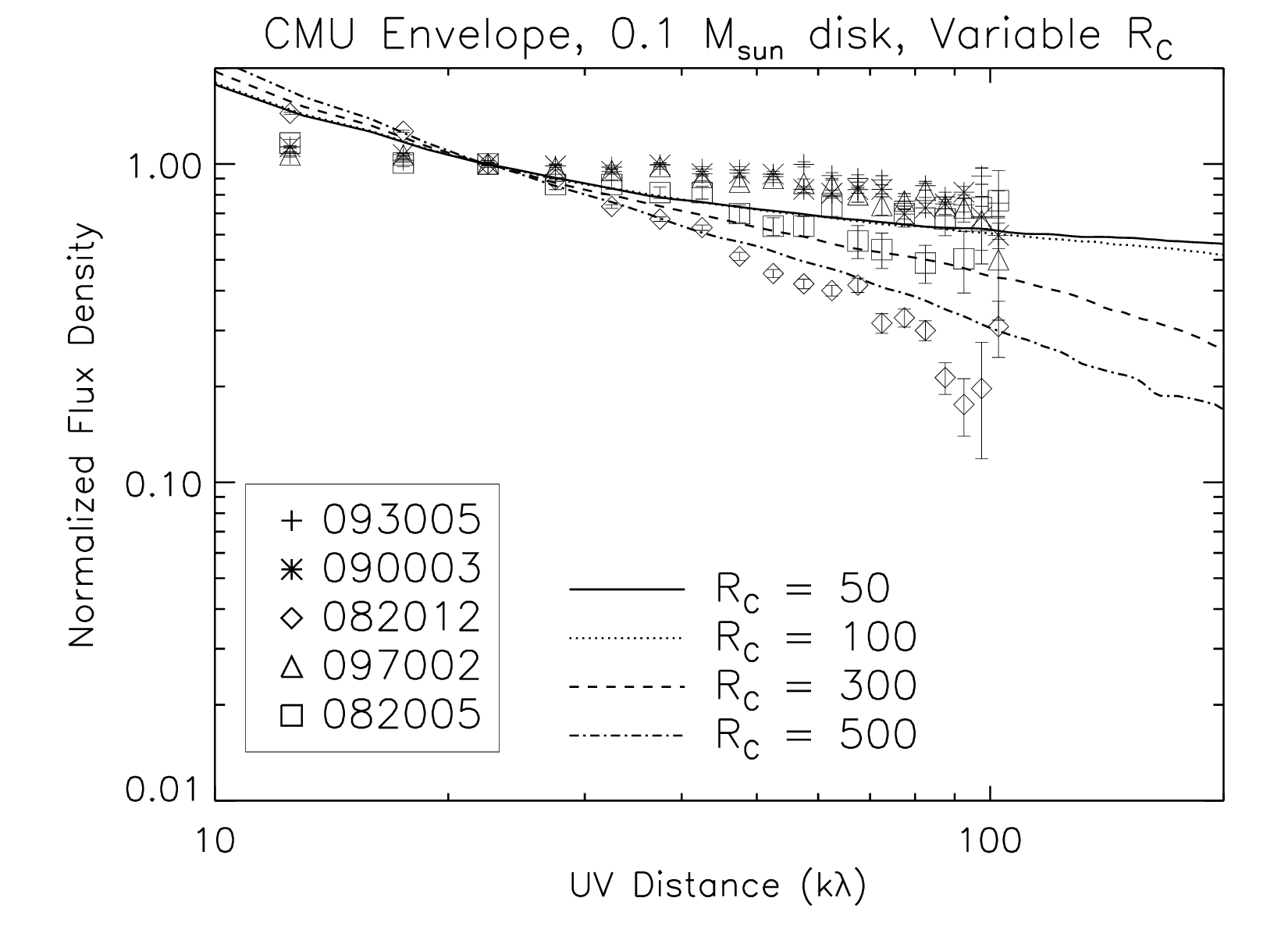}
\end{center}
\caption{Same as Figure \ref{uvcomps}, but for CMU, rotationally flattened envelopes with various centrifugal radii.}
\label{uvcomps-cmu}
\end{figure}

\begin{figure}
\begin{center}

\includegraphics[scale=0.66]{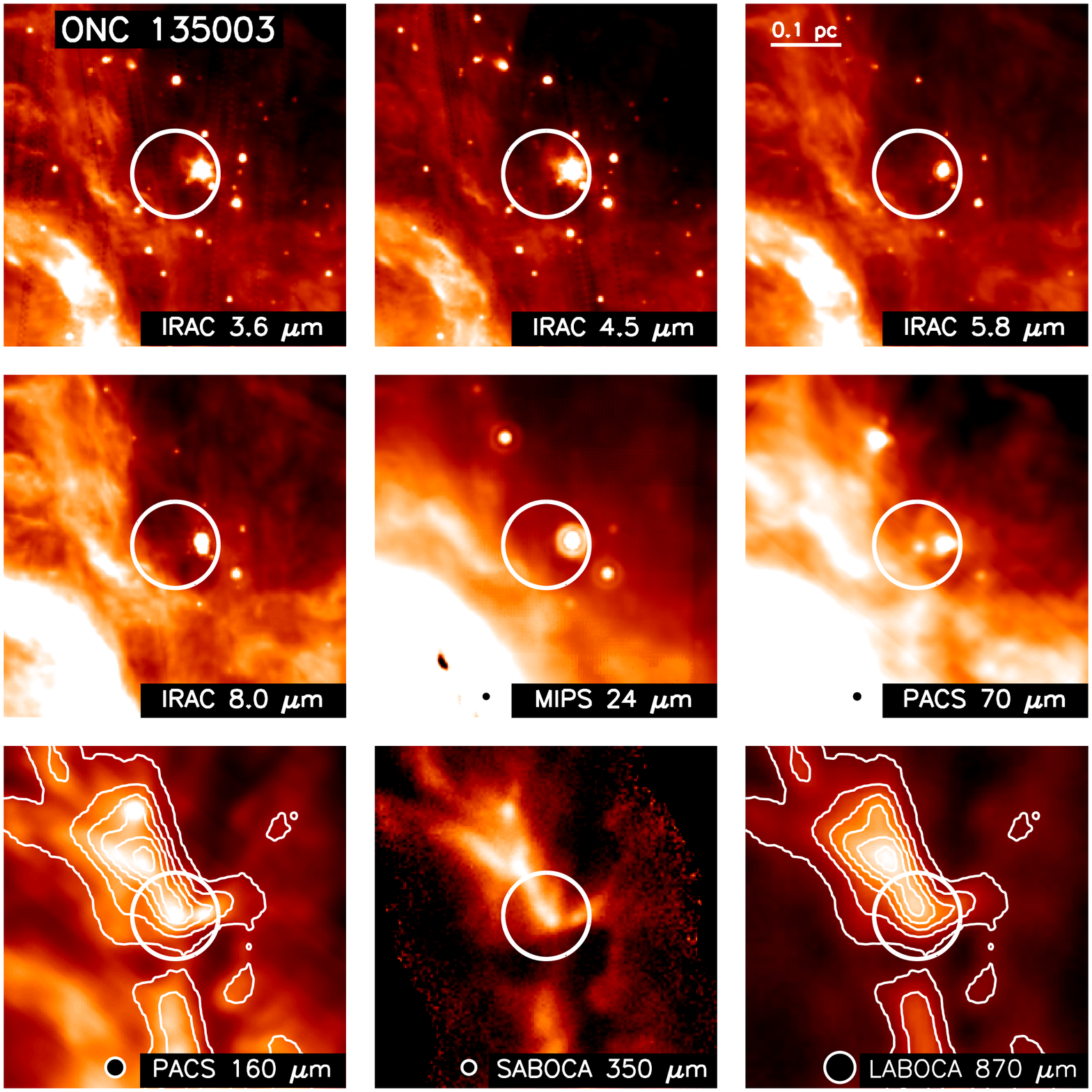}

\end{center}
\caption{Multi-wavelength images of the additional PBRS source 135003. This source is located in the OMC 2/3 region 
where there are high levels of nebulosity. The source immediately west of 135003 is HOPS 59, a Class I protostar. Notice
that 135003 resides at an inflection point of the filament traced at longer wavelengths. The white circle marks the location of 135003 and is
30\arcsec\ in radius. The black circles at the bottom are the respective beam sizes. 
The contours overlaid on the 160 \micron\ and 870 \micron\ 
images are the 870 \micron\ emission contours at [0.5, 1.0, 1.5, 2.0, 2.5, and 3.0] Jy beam$^{-1}$.
}
\label{135003}
\end{figure}

\begin{figure}
\begin{center}

\includegraphics[scale=1.0]{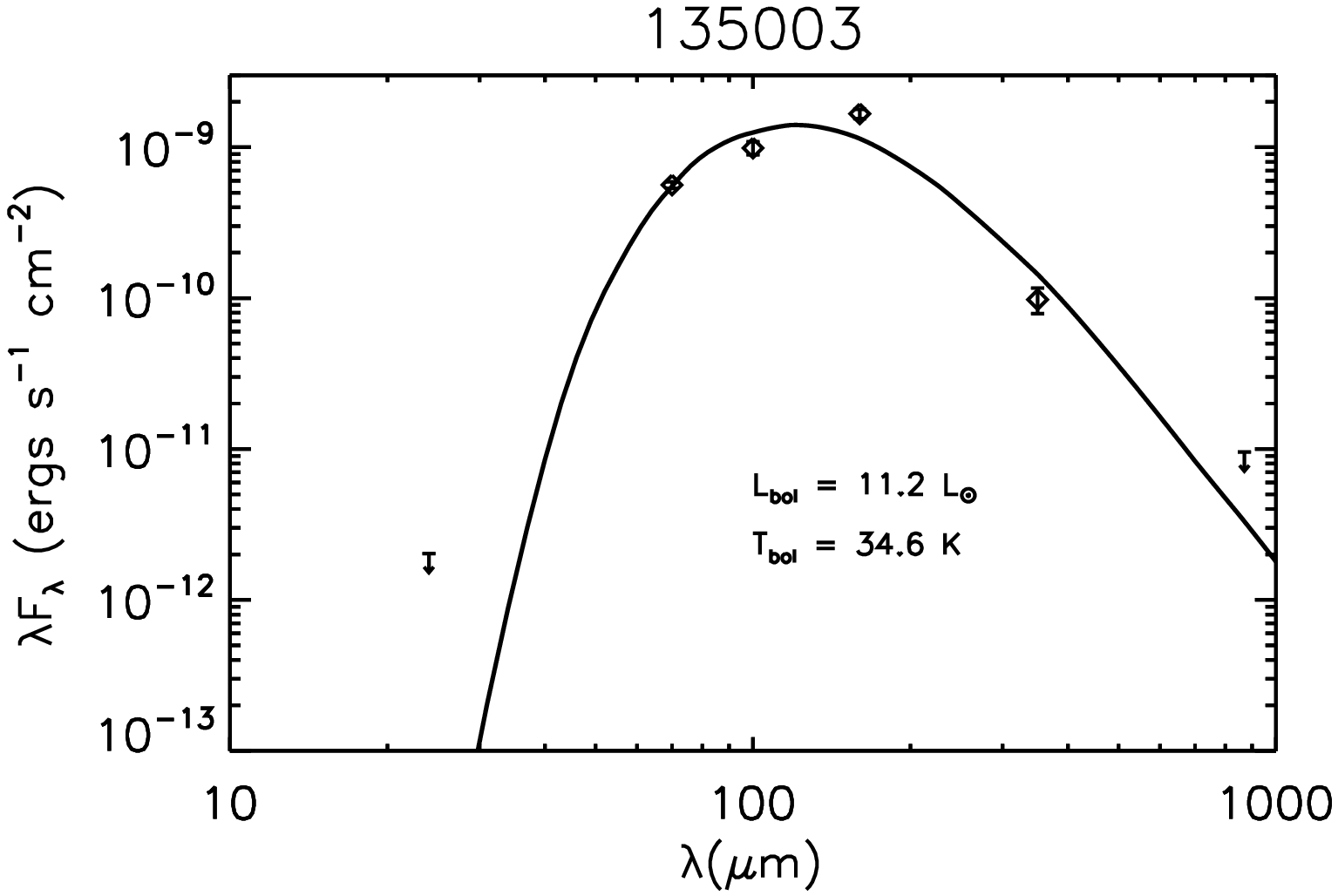}

\end{center}
\caption{SED of 135003 including data from \textit{Spitzer}, \textit{Herschel}, and APEX. The APEX 870 \micron\
point is an upper limit due to source blending.}
\label{sed-135003}
\end{figure}

\clearpage
\begin{deluxetable}{llcl}
\tablewidth{0pt}
\tabletypesize{\scriptsize}
\tablecaption{CARMA 2.9 mm Observation Log}
\tablehead{
  \colhead{Source(s)}   & \colhead{Date} & \colhead{Configuration} & \colhead{Calibrators}\\
               &    \colhead{(UT)}      &    &   \colhead{(Gain, Flux)}   \\
}
\startdata
093005, 091015        & 04 Oct 2012 & D & 0532+075, Uranus      \\
090003, 082012\tablenotemark{a}, 302002  & 10 Oct 2012 & D & 0532+075, None \\
093005, HOPS373                          & 12 Oct 2012 & D & 0532+075, Uranus      \\
061012\tablenotemark{b}, 082005          & 13 Oct 2012 &D &  0532+075, Uranus      \\
090003, 082012\tablenotemark{a}, 302002  & 15 Oct 2012 &D &  0532+075, Jupiter \\
061012\tablenotemark{b}, 082005          & 17 Oct 2012 & D & 0532+075, Uranus      \\
119019, 097002                           & 17 Jan 2014 & D & 0607-085, 3C84      \\
119019, 097002                           & 31 Jan 2014 & D & 0607-085, 3C84      \\
135003, 019003                           & 13 Feb 2014 & D & 0607-085, Uranus      \\
135003, 019003                           & 20 Feb 2014 & D & 0607-085, Uranus      \\

093005, 090003, HOPS 373                 & 09 Mar 2014 & C & 0607-085, Uranus      \\
093005, 090003, HOPS 373                 & 13 Mar 2014 & C & 0607-085, 3C84      \\
082005, 097002, 082012\tablenotemark{a}  & 07 Mar 2014 & C & 0607-085, 3C84      \\
082005, 097002, 082012\tablenotemark{a}  & 14 Mar 2014 & C & 0607-085, Uranus      \\
091015, 091016                           & 16 Mar 2014 & C & 0607-085, Uranus      \\
091015, 091016                           & 18 Mar 2014 & C & 0607-085, Uranus      \\
091016\tablenotemark{c}                                   & 04 May 2014 & D & 0607-085, 3C84, 3C273      \\

%\\
\enddata
\tablecomments{0423-013 was used as the bandpass calibrator for all observations in 2012. The bandpass calibrator
was 3C84 for all later observations.} 
\tablenotetext{a}{HOPS 372 is also in the field of 082012.}
\tablenotetext{b}{HOPS 223 is also in the field of 061012.}
\tablenotetext{c}{Data obtained through the CARMA FastTrack System.}
\end{deluxetable}

\begin{deluxetable}{llllllccllll}
\tablewidth{0pt}
\rotate
\tabletypesize{\scriptsize}
\tablecaption{2.9 mm Continuum Fluxes}
\tablehead{
  \colhead{Source} & \colhead{PBRS} & \colhead{HOPS ID} & \colhead{RA} & \colhead{Dec} & \colhead{$S_{\nu}$} & \colhead{Peak} & \colhead{Noise} & \colhead{Beam} & \colhead{Mass} & \colhead{L$_{bol}$} & \colhead{T$_{bol}$}\\
             &         \colhead{(yes/no)}              & & \colhead{(J2000)} &  \colhead{(J2000)}  & \colhead{(mJy)} & \colhead{(mJy beam$^{-1}$)} & \colhead{(mJy beam$^{-1}$)} & \colhead {(\arcsec)} & \colhead{($M_{\sun}$)} & \colhead{(L$_{\sun}$)} & \colhead{(K)}\\
}
\startdata
097002   & yes & 404 & 05:48:07.71 & +00:33:51.7 &  45.7  $\pm$ 2.5 & 39.0 & 0.4  & 3.31 $\times$ 2.74 & 2.8 $\pm$ 0.3 & 1.14 & 33.4\\
HOPS 373 & yes & 373 & 05:46:30.99 & -00:02:33.9 &  51.1  $\pm$ 4.2 & 17.0 & 0.45 & 2.08 $\times$ 1.97 & 3.1 $\pm$ 0.4 & 5.2 & 36.0\\
302002   & yes & 407 & 05:46:28.28 & +00:19:28.1 &  47.1  $\pm$ 2.5 & 27.0 & 0.6  & 5.00 $\times$ 4.03 & 2.9 $\pm$ 0.3 & 0.85 & 28.6\\
093005   & yes & 403 & 05:46:27.90 & -00:00:52.1 &  90.0  $\pm$ 4.8 & 72.2 & 0.6  & 2.42 $\times$ 2.29 & 5.4 $\pm$ 0.6 & 1.7 & 30.8\\
091016\tablenotemark{a} & yes & 402 & 05:46:10.01 & -00:12:17.3 &  46.6  $\pm$ 2.7 & 42.1 & 0.6  & 2.15 $\times$ 2.09 & 2.8 $\pm$ 0.3 & 0.65 & 29.1\\
091015\tablenotemark{a} & yes &  401 & 05:46:07.72 & -00:12:21.3 &  30.9  $\pm$ 2.6 & 17.4 & 0.7  & 2.19 $\times$ 2.12 & 1.9 $\pm$ 0.3 & 0.81 & 30.9\\
061012   & yes & 397  & 05:42:49.03 &  -08:16:11.8 &  17.0  $\pm$ 3.1 & 6.5  & 0.8  & 5.58 $\times$ 4.25 & 1.0 $\pm$ 0.2 & 0.75 & 32.1\\
HOPS 223 & no  & 223 & 05:42:48.47 & -08:16:34.3 &  49.6  $\pm$ 3.9 & 28.2 & 1.0 & 5.58 $\times$ 4.25 & 3.0 $\pm$ 0.4 & 28.0 & 136.0\\
090003   & yes & 400 & 05:42:45.26 &  -01:16:13.9 &  115.4 $\pm$ 3.9 & 83.3 & 0.55 & 2.74 $\times$ 2.56 & 7.0 $\pm$ 0.7 & 2.71 & 36.0\\
082005   & yes & 398 & 05:41:29.40 &  -02:21:16.5 &  32.9  $\pm$ 3.0 & 21.3 & 0.4  & 2.61 $\times$ 2.4  & 2.0 $\pm$ 0.3 & 1.02 & 29.3\\
HOPS 372 & yes & 372 & 05:41:26.34 &  -02:18:21.6 &  35.6  $\pm$ 5.4 & 10.2 & 0.7  & 2.52 $\times$ 2.38 & 2.2 $\pm$ 0.4 & 4.9 & 36.9\\
082012   & yes & 399 & 05:41:24.92 &  -02:18:07.0 &  155.6 $\pm$ 4.6 & 87.5 & 0.6  & 2.52 $\times$ 2.38 & 9.4 $\pm$ 1.0 & 6.3 & 32.2\\
119019   & yes & 405 & 05:40:58.56 &  -08:05:35.0 &  10.2   $\pm$ 1.8 & 4.0  & 0.5  & 6.89 $\times$ 3.71 & 0.6 $\pm$ 0.1 & 1.56 & 34.4\\
019003 A\tablenotemark{b} & yes &  394 & 05:35:24.23 & -05:07:53.9 &  38.9 $\pm$ 0.9  & 16.8  & 1.1 & 5.58 $\times$ 4.12 & 2.4 $\pm$ 0.3 & 3.16 & 33.6\\
019003 B\tablenotemark{b} & no &      & 05:35:24.86 & -05:07:53.4 &  52.7 $\pm$ 0.9 & 24.3  & 1.1 & 5.58 $\times$ 4.12 & 3.2 $\pm$ 0.3 & $<$ 0.5 & $<$20.0\\
HOPS 68 & no & 68  & 05:35:24.27 & -05:08:32.2 &  41.8 $\pm$ 5.9 & 21.2  & 1.8  & 6.91 $\times$ 4.8  & 2.5 $\pm$ 0.4 & 5.7 &  92.9\\
HOPS 71\tablenotemark{a} & no & 71 & 05:35:23.19 & -05:07:45.1 &  5.6   $\pm$ 2.2 & 9.4  & 1.1  & 5.58 $\times$ 4.12 & 0.3 $\pm$ 0.1 & 6.4 & 231.0\\
135003  & yes & 409 & 05:35:21.40  & -05:13:17.5 &  49.3  $\pm$ 3.9 & 16.9 & 1.0 & 5.53 $\times$ 4.14 & 3.0 $\pm$ 0.4 & 12.0 & 30.0\\
HOPS 59 & no & 59 & 05:35:20.15 & -05:13:15.7 &  40.4  $\pm$ 6.5 & 29.5 & 1.1 & 5.53 $\times$ 4.14 & 2.4 $\pm$ 0.5 & 49.5 &  479.0 \\
\enddata
\tablecomments{Unless otherwise noted, the flux densities
 were measured in 20\arcsec\ $\times$ 20\arcsec\ regions centered on the 
protostars and the positions were determined from Gaussian fits to the 2.9 mm data. The uncertainty of the integrated
flux density is calculated from the uncertainty in the summation of all flux within the region; 
$\sigma_{int}$ = N$^{1/2}$ $\sigma$, where
N is the number of independent beams within the integrated region and $\sigma$ is the rms noise. } 
\tablenotetext{a}{Flux densities were measured in 10\arcsec\ $\times$ 10\arcsec\ regions to avoid neighboring sources and negatives.}
\tablenotetext{b}{Flux densities were derived from Gaussian fitting due to blended sources.}
\end{deluxetable}

\begin{deluxetable}{lllll}

\tablewidth{0pt}
\tabletypesize{\scriptsize}
\tablecaption{Literature Flux Densities}
\tablehead{
  \colhead{Source} & \colhead{$S_{\nu}$} & \colhead{Distance} & \colhead{L$_{bol}$} & \colhead{$\lambda$}\\
               & \colhead{(mJy)}         &\colhead{(pc)}  & \colhead{(L$_{\sun}$)}   &\colhead{(mm)}
}
\startdata
L1448 IRS3B    & 134.6 $\pm$ 3.9  & 230 & 5.0 & 2.7\\
L1448 IRS3C    & 31.7 $\pm$ 4.1   & 230 & 1.0 & 2.7\\
NGC 1333 IRAS2A & 82.8 $\pm$ 4.0   & 230 & 20.0  & 2.7\\
NGC 1333 IRAS4A & 544.2 $\pm$ 13.6 & 230 & 5.8 & 2.7\\
NGC 1333 IRAS4B1& 180.3 $\pm$ 7.9  & 230 & 3.8 & 2.7\\
NGC 1333 IRAS4B2& 49.8 $\pm$ 5.5   & 230 & 3.8 & 2.7\\
L1551 IRS5     & 173.3 $\pm$ 7.5  & 140 & 24.5  & 2.7\\
VLA 1623       & 72.1 $\pm$ 6.8   & 125 & 1.0 & 2.7\\
IRAS 16293-2422& 1017.9 $\pm$ 26.5& 125 & 15.0 & 2.7\\
\\
Perseus 5       & 4.0 $\pm$ 1.0  & 230 & 0.5 & 3.4\\
HH211           & 43.4 $\pm$ 2.0 & 230 & 1.2 & 3.4\\
L1527           & 32.6 $\pm$ 6.0 & 140 & 1.7 & 3.4 \\
HH270VLA1       & 6.3 $\pm$ 1.0  & 420 & 7.0 & 3.4\\
RNO43           & 9.4 $\pm$ 3.5  & 420 & 12.5 & 3.4\\
IRAS 16253-2429 & 1.8 $\pm$ 1.0  & 125 & 0.25 & 3.4\\
HH108MMS        & 10.3 $\pm$ 2.3 & 300 & 0.7 & 3.4\\
HH108IRS        & 35.7 $\pm$ 4.9 & 300 & 8.0 & 3.4\\
L1165           & 12.1 $\pm$ 1.7 & 300 & 13.0 & 3.4\\
L1152           & 15.2 $\pm$ 2.8 & 300 & 1.0 & 3.4\\
L1157-CARMA     & 55.1 $\pm$ 8.3 & 300 & 5.0 & 3.4\\
\\
RNO43         & 26  $\pm$  6   &  420  &   12.5  & 2.7\\
HH114MMS       &110 $\pm$  20 &   420  &   26.0  & 2.7\\
IRAS03282+3035 &61  $\pm$  12 &   230  &   1.2  & 2.7\\
IRAS04239+2436& 14  $\pm$  2   &  140  &   1.1  & 2.7\\
IRAS20582+7724& 45 $\pm$   9   &  200  &   4.0  & 2.7\\
IRAS21004+7811 &62 $\pm$   13  &  200  &   1.0  & 2.7\\

\enddata
\tablecomments{The collection of 2.7 mm data at the top are from \citet{looney2000}, the 3.4 mm data are from \citet{tobin2011}, and
the collection of 2.7 mm data at the bottom are from \citep{arce2006}. } 

\end{deluxetable}

\begin{deluxetable}{ll}

\tablewidth{0pt}
\tabletypesize{\scriptsize}
\tablecaption{135003 Photometry}
\tablehead{
\colhead{Wavelength} & \colhead{Flux Density}\\
\colhead{(\micron)}         &\colhead{(Jy)}
}
\startdata
   24.0 &  $<$ 0.016\\
   70.0 &  13.1 $\pm$ 0.7\\
  100.0 &  32.9 $\pm$ 3.3\\
  160.0 &  88.9 $\pm$ 6.2\\
  350.0 &  11.4 $\pm$ 2.2\\
  870.0 &  $<$ 2.8\\

\enddata

\end{deluxetable}

\end{document}